\documentclass{article}

\usepackage{arxiv}

\usepackage[utf8]{inputenc}
\usepackage[T1]{fontenc}
\usepackage[english]{babel}
\usepackage[hidelinks]{hyperref}
\usepackage{url}
\usepackage{graphicx}
\usepackage{booktabs}
\usepackage{amsmath, amssymb}
\usepackage{microtype}
\usepackage{placeins}

\usepackage{natbib}
\title{A Retrospective Benchmark of Spatiotemporal Covariates for Daily Active-Fire Detection in Cerrado Conservation Units}

\author{
  Juliano Eleno Silva P\'adua\thanks{\texttt{julianopadua@estudante.ufscar.br}},
  Alexandre Luis Magalh\~aes Levada\thanks{\texttt{alexandre.levada@ufscar.br}},
  Fredy Joao Valente\thanks{\texttt{fredy@ufscar.br}} \\
  Computing Department\\
  Federal University of S\~ao Carlos\\
  13565-905, S\~ao Carlos-SP, Brazil
}

\begin{document}

\maketitle

\begin{abstract}
Wildfires threaten biodiversity, carbon stocks, and management capacity in the Brazilian Cerrado, where Conservation Units (CUs) and their official buffer zones (Zonas de Amortecimento) must allocate prevention resources under a strong dry-season fire regime.
Operational systems published by the Instituto Nacional de Pesquisas Espaciais (INPE) issue continental dryness indices and reactive hotspot alerts, but neither output provides a CU-scale benchmark for ranking daily active-fire detections.
This work develops a retrospective daily active-fire detection benchmark for the Cerrado portion of Minas Gerais, Brazil, using INPE BDQueimadas reference satellite labels (AQUA\_M-T), rule-bound pseudo absences sampled inside the Cerrado-MG mask with same-year MapBiomas Collection 9 land-cover filtering, and a four-stage progressive feature ablation extracted through Google Earth Engine.
A model triad, Logistic Regression, Random Forest, and XGBoost, is evaluated under five-fold time series cross validation on a global training base and on an independent imbalanced test set spatially held out to Parque Estadual do Pau Furado and Parque Estadual da Serra do Cabral with their official buffer zones.
AUC-PR is the primary metric, with AUC-ROC, threshold precision and recall, Shapley Additive Explanations (SHAP), and retrospective score maps used as complementary diagnostics.
Temporal cross-validation in the Cerrado-MG training domain showed the highest mean AUC-PR at the complete temporal-memory stage for all three model families.
Held-out AOI tests were weaker: absolute AUC-PR decreased under the stricter 1:100 prevalence design, Random Forest peaked at Stage 3 in both AOIs, and XGBoost maps exposed high-recall, high-warning-volume behavior.
The resulting baseline provides a reproducible reference for comparing atmospheric, surface, static spatial, and short-term memory covariates in daily CU-scale active-fire detection ranking.
Because Stages 1 to 3 use same-day covariates, it is not a prospective forecast; that extension requires predictors available before the decision time.

\end{abstract}

\section{Introduction}
The Cerrado is a fire dependent savanna biome that has coevolved with seasonal burning, and where ecological fire history, dry season fuel curing, anthropogenic ignition, and accelerating land use pressure converge~\citep{hardesty2005fire,pivello2011use,dieleviegas2022brazilianbiomes}.
According to the Cadastro Nacional de Unidades de Conservação (CNUC), the Cerrado biome currently shelters about 177,719 km$^2$ of protected area distributed across 438 Conservation Units (CUs), of which 142 are under integral protection~\citep{hoffmann2020fire}.
Inside those CUs, fire management is a daily allocation problem in which managers must decide when and where patrols, prevention activities, and suppression readiness should be intensified.
\citet{hoffmann2020fire} document that the percentage of burned area inside Cerrado CUs and their official buffer zones varies sharply with adjacent land cover and dry season rainfall anomalies, which makes the inside versus buffer distinction operationally consequential.

The Brazilian operational baseline for fire hazard is the Risco de Fogo (RF) index, issued daily on a continental 10 km grid by the Instituto Nacional de Pesquisas Espaciais (INPE) Programa Queimadas~\citep{setzer2019riscofogo}.
The RF index folds 120 days of accumulated rainfall, a sinusoidal vegetation specific dryness ramp, a humidity factor, a temperature factor, and a latitude and elevation correction into a dimensionless score; it explicitly omits wind, soil moisture, and statistical uncertainty quantification~\citep{setzer2019riscofogo}.
At CU scale, where the typical Cerrado state park is much smaller than a single 10 km RF pixel, the index can only describe the regional envelope, not an AOI-specific active-fire detection score with quantified uncertainty.
The complementary operational tool, INPE BDQueimadas, reports active fire detections \emph{after} thermal anomalies have been observed by the reference satellite AQUA\_M-T~\citep{inpebdq}.
What remains missing is a CU-scale benchmark that tests whether environmental and spatial covariates can rank daily BDQueimadas active-fire detection occurrence under realistic class imbalance, while also identifying what would still be needed for future prospective forecasting.

Machine learning has repeatedly achieved competitive accuracy for regional wildfire occurrence and susceptibility modeling, with reported area under the receiver operating characteristic curve (AUC-ROC) values commonly between 0.80 and 0.95~\citep{jain2020review,cilli2022xai,yang2026forests,bian2024forests,freitas2025triunfo}.
Recent studies increasingly pair those models with explainable artificial intelligence (XAI), especially Shapley Additive Explanations (SHAP), to inspect fitted behavior in terms of physically interpretable drivers~\citep{gunning2019xai,guidotti2018survey,lundberg2017shap}.
\citet{yang2026forests} combine XGBoost and SHAP on a kilometer grid in Yunnan, China; \citet{cilli2022xai} combine Random Forest, SHAP, and hotspot statistics for decade aggregated Mediterranean fire occurrence; \citet{bian2024forests} compare Logistic Regression, Random Forest, and XGBoost on Chinese provincial fire data; and \citet{freitas2025triunfo} provide the closest Brazilian protected-area precedent, using Random Forest, XGBoost, SHAP, BDQueimadas, MapBiomas, CHIRPS, SRTM, and MODIS products in the Triunfo do Xingu Amazonian Environmental Protection Area.
These studies establish the usefulness of machine-learning and XAI tools for fire susceptibility, but they do not provide a daily CU-scale Cerrado benchmark with rare-class AOI hold-out testing and row-aligned feature ablation.
\citet{phelps2021guidelines} further caution that balanced datasets and AUC-ROC-centered validation can mask the false positive cost that dominates operational deployment under severe class imbalance.

This paper evaluates spatiotemporal covariates for \emph{daily} active-fire detection occurrence inside the Cerrado portion of Minas Gerais, Brazil, and quantifies the marginal value of four progressively richer feature families for retrospective classification and ranking.
We compare a model triad of Logistic Regression as a linear reference, Random Forest as a bagging baseline, and XGBoost as a boosting baseline, under time series cross validation on a global training base, plus an independent imbalanced area of interest (AOI) test set held out to two state CUs and their buffer zones: Parque Estadual do Pau Furado in Uberlândia, Minas Gerais (CNUC 0000.31.1770) and Parque Estadual da Serra do Cabral in Augusto de Lima, Minas Gerais (CNUC 0000.31.0890).
The four feature stages, each a strict superset of the previous one, are: Stage 1 atmospheric reanalysis from the European Centre for Medium Range Weather Forecasts (ECMWF) Reanalysis v5 Land (ERA5-Land); Stage 2 optical and thermal surface state from MODIS; Stage 3 static topography and built environment proximity; Stage 4 light causal exponentially weighted memory over selected Stage 1 and Stage 2 inputs.
AUC-PR is the primary metric, with AUC-ROC, precision, and recall retained as complementary diagnostics for the explicitly imbalanced AOI test set~\citep{phelps2021guidelines}.
To support external reproduction, a public companion repository with the pipeline scaffold is available online.\footnote{\url{https://github.com/julianopadua/cerrado-wildfire-prediction-reproducibility}}

The paper makes three contributions:
\begin{enumerate}
  \item It builds a retrospective daily active-fire detection baseline for the Cerrado portion of Minas Gerais using BDQueimadas positives, constrained pseudo absences, and an independent AOI test protocol for two held-out CUs and their buffer zones.
  \item It compares Logistic Regression, Random Forest, and XGBoost under identical row-aligned feature matrices across four nested spatiotemporal ablation stages.
  \item It quantifies the marginal value of atmospheric, surface, static spatial, and short-term temporal memory covariates with AUC-PR-centered validation, SHAP diagnostics, and retrospective operational-style score maps.
\end{enumerate}

Section~\ref{sec:theory} defines the modeling background.
Section~\ref{sec:data} describes the data sources, label processing, pseudo absences, AOI tests, and covariate stages.
Section~\ref{sec:methods} presents ablation, tuning, validation, AOI testing, and retrospective mapping.
The Results and Discussion sections report and interpret the evaluation.

\section{Theoretical background}

\label{sec:theory}

This section defines the modeling elements used in the daily active-fire detection study: rare-event classification, pseudo absence design, the three classifier families, the covariate stages, temporal memory features, evaluation metrics, and SHAP explanations.

\subsection{Daily fire occurrence as a rare event problem}

For each sampled point-date tuple $i$, the response variable is
\begin{equation}
  y_i =
  \begin{cases}
    1, & \text{if a retained BDQueimadas active fire detection is observed,}\\
    0, & \text{if the tuple is sampled as a constrained pseudo absence.}
  \end{cases}
\end{equation}
The modeling objective is to estimate a score for $p(y_i=1 \mid \mathbf{x}_i)$, where $\mathbf{x}_i$ contains atmospheric, surface, static spatial, seasonal, and temporal memory covariates extracted for the same point and date.
Because true non fire events are not directly observed, negative samples are pseudo absences rather than confirmed absence records.
The pseudo absence design defines which background point-dates are treated as plausible non-fire examples under the same observation process.

\citet{barbetmassin2012pseudo} describe pseudo absence design through three questions: how background samples are generated, where they are allowed, and how many are drawn relative to presences.
Naive background points can make fire classification artificially easy by drawing non-fire samples from irrelevant land covers, seasons, or held-out evaluation geographies.
Our design constrains background samples through the Cerrado within Minas Gerais mask, dynamic MapBiomas land cover, same-day distance from positives, dry-season timing, and the spatial hold-out of target CUs and buffer zones.
This differs from density-regression studies such as \citet{freitas2025triunfo}, where the response is a smoothed fire-density surface and the pseudo absence question is replaced by a smoothing-radius question.

Under deployment-like daily prevalence, active-fire detection ranking is severely imbalanced.
\citet{phelps2021guidelines} show that wildfire occurrence models can appear similar under balanced evaluation but diverge sharply when tested at deployment-like prevalence, especially when AUC-ROC is treated as the main metric.
The design therefore separates the training prevalence, controlled through pseudo absence ratios, from the independent area of interest (AOI) test prevalence, which is intentionally rarer and closer to the protected-area decision setting.

\subsection{Classifier families}

Logistic Regression is used as the linear reference model.
For covariates $\mathbf{x}_i \in \mathbb{R}^p$, it models the log odds of active-fire detection occurrence as
\begin{equation}
  \log\left(\frac{p_i}{1-p_i}\right) = \beta_0 + \mathbf{x}_i^\top \boldsymbol{\beta},
  \qquad
  p_i = \frac{1}{1 + \exp[-(\beta_0 + \mathbf{x}_i^\top \boldsymbol{\beta})]}.
\end{equation}
Its coefficient signs and magnitudes provide a transparent baseline for whether the feature stages contain a mostly linear fire signal~\citep{hosmer2013appliedlogistic}.

Random Forest is the bagging baseline~\citep{breiman2001random}.
It trains an ensemble of decision trees on bootstrap samples and random subsets of candidate predictors.
For classification, the estimated probability score is the average vote across trees,
\begin{equation}
  \hat{p}_i = \frac{1}{T}\sum_{t=1}^{T} h_t(\mathbf{x}_i),
\end{equation}
where $h_t$ is the probability output or vote returned by tree $t$.
Random Forest is widely used in wildfire mapping because it handles nonlinear interactions, mixed feature scales, and threshold effects without strong parametric assumptions~\citep{cilli2022xai,jain2020review}.

XGBoost is the boosting baseline and the highest-capacity member of the triad~\citep{chen2016xgboost}.
It represents the prediction as an additive tree ensemble,
\begin{equation}
  \hat{y}_i = \sum_{k=1}^{K} f_k(\mathbf{x}_i), \qquad f_k \in \mathcal{F},
\end{equation}
and minimizes a regularized objective,
\begin{equation}
  \mathcal{L} = \sum_i \ell(y_i,\hat{y}_i) + \sum_{k=1}^{K} \Omega(f_k),
\end{equation}
where $\ell$ is the classification loss and $\Omega$ penalizes tree complexity.
Recent wildfire susceptibility studies report strong XGBoost performance with climate, vegetation, terrain, and anthropogenic predictors~\citep{yang2026forests,bian2024forests,freitas2025triunfo}.
Here, XGBoost is tested against a linear reference and a bagged-tree baseline under the same row-aligned feature stages.

\subsection{Spatiotemporal covariates and environmental memory}

The covariate design represents four fire-risk dimensions: atmospheric dryness, surface condition and fuel water stress, terrain and built-environment context, and recent environmental history~\citep{jain2020review,setzer2019riscofogo,yang2026forests,bian2024forests}.

Relative humidity is derived from air temperature and dewpoint temperature using the Tetens form also reported in \citet{yang2026forests}.
Let $T$ be two meter air temperature and $T_d$ be two meter dewpoint temperature in degrees Celsius.
The saturation and actual vapor pressures are
\begin{equation}
  e_s(T) = 0.6108 \exp\left(\frac{17.27T}{T + 237.3}\right),
  \qquad
  e(T_d) = 0.6108 \exp\left(\frac{17.27T_d}{T_d + 237.3}\right),
\end{equation}
and relative humidity is
\begin{equation}
  RH = 100 \frac{e(T_d)}{e_s(T)}.
\end{equation}
Wind is represented through speed and circular direction components derived from the ERA5-Land zonal and meridional wind bands, avoiding a discontinuity at 0 and 360 degrees.

Temporal memory is represented with causal exponentially weighted moving averages (EWMAs).
For a variable $z_t$ and smoothing factor $\alpha$, the recursive form is
\begin{equation}
  m_t = \alpha z_t + (1-\alpha)m_{t-1},
\end{equation}
with $\alpha = 1 - 2^{-1/h}$ for half life $h$ days.
The construction is causal because only observations at or before the target date enter the average.
This choice follows antecedent-rainfall logic, where previous rainfall is summarized with temporally decaying weights that emphasize recent precipitation and serve as proxies for antecedent wetness or soil moisture.
Fire-danger systems also encode environmental memory through multiple fuel-moisture response times: the Canadian Fire Weather Index System combines fast fine-fuel, intermediate duff, and slower drought components, while the Brazilian Risco de Fogo index carries a weighted rainfall history up to 120 days~\citep{vanwagner1987fwi,setzer2019riscofogo}.
Here, the EWMA design extends this rainfall-history principle to heat, humidity, precipitation, greenness, water stress, and land surface temperature without tuning the lag structure separately for each classifier.

\subsection{Evaluation and explanation}

Temporal validation is required because random splits can leak nearby dates from the same dry season into both training and validation.
Ordered time-series folds are used for training evaluation, and the two target CUs with their official buffer zones are reserved for an independent AOI test.
The AOI test reuses the same point set across all feature stages, so differences across stages measure marginal feature value rather than a changing test composition.

Precision and recall are defined as
\begin{equation}
  \mathrm{precision} = \frac{TP}{TP + FP},
  \qquad
  \mathrm{recall} = \frac{TP}{TP + FN}.
\end{equation}
These definitions underlie the AUC-PR, AUC-ROC, and threshold diagnostics reported later.
The metric hierarchy is specified in Section~\ref{sec:methods}, where AUC-PR is used as the primary rare-event ranking metric and AUC-ROC remains a secondary discrimination check~\citep{davis2006prroc,saito2015precision,phelps2021guidelines}.
Precision and recall at a fixed threshold expose the decision balance between false alarms and missed detections.
F1 score, confusion matrices, and calibration scores such as the Brier score are better reserved for a threshold-selection and calibration analysis once an operational decision rule is specified.

Finally, the explanation layer uses Shapley Additive Explanations (SHAP) to decompose model predictions into additive feature contributions~\citep{lundberg2017shap,gunning2019xai,guidotti2018survey}.
For an instance $\mathbf{x}$, SHAP writes the model output as
\begin{equation}
  g(\mathbf{x}) = \phi_0 + \sum_{j=1}^{p}\phi_j,
\end{equation}
where $\phi_0$ is the expected prediction and $\phi_j$ is the contribution assigned to feature $j$.
The explanation layer makes the learned daily active-fire scores inspectable and comparable with the physical assumptions embedded in operational indices and previous wildfire susceptibility studies~\citep{cilli2022xai,yang2026forests,freitas2025triunfo}.

\section{Data and pre-processing}

\label{sec:data}

The dataset combines INPE BDQueimadas active fire detections, Cerrado and administrative boundaries, annual land-cover rasters, reanalysis weather fields, MODIS optical and thermal products, elevation, and built-environment proximity layers.
Each data family is described by its source, filtering operation, spatial or temporal support, and role in the modeling table.

\subsection{Study area}

\begin{figure}[!htbp]
  \centering
  \includegraphics[width=0.76\linewidth]{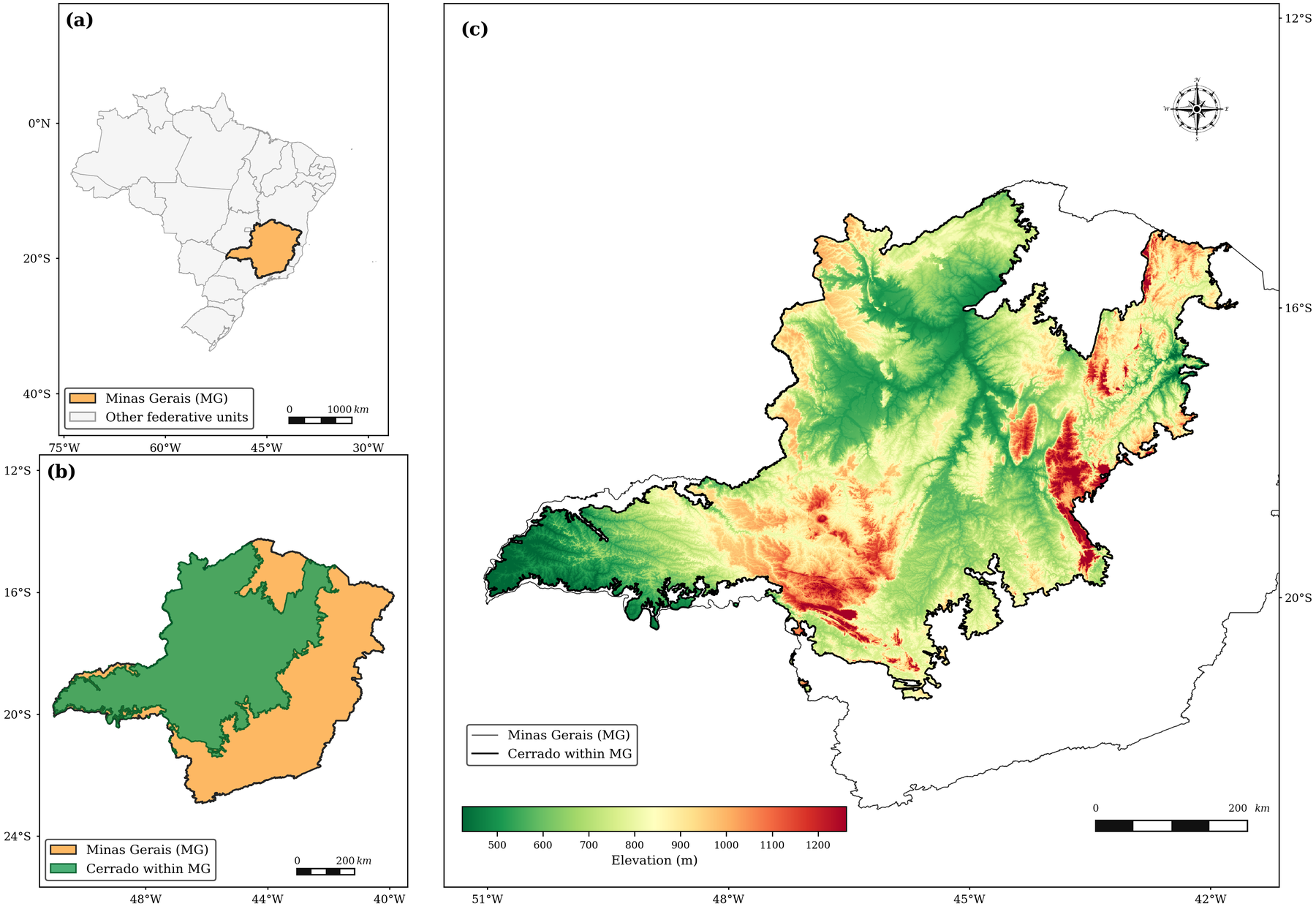}
  \caption{Cerrado-MG study area. Panel (\textbf{a}) shows Minas Gerais in Brazil; panel (\textbf{b}) shows the Cerrado footprint inside Minas Gerais; panel (\textbf{c}) shows Copernicus DEM clipped to the extraction mask.}
  \label{fig:study-area}
\end{figure}

The modeling footprint is the Cerrado biome restricted to Minas Gerais, Brazil.
It was obtained by intersecting the IBGE Minas Gerais state polygon with the IBGE Cerrado biome polygon under the South American Geocentric Reference Frame (SIRGAS) 2000 datum, using boundaries retrieved through the \texttt{geobr} package~\citep{pereira2026geobr}.
This Cerrado-MG mask defines the spatial extent for the global training base, the land-cover diagnostics, and the raster extractions.
Figure~\ref{fig:study-area} summarizes this hierarchical spatial context, from Brazil to Minas Gerais and then to the Cerrado-MG extraction mask.

\FloatBarrier

\begin{figure}[!htbp]
  \centering
  \begin{minipage}[t]{0.94\linewidth}
    \centering
    \includegraphics[width=\linewidth]{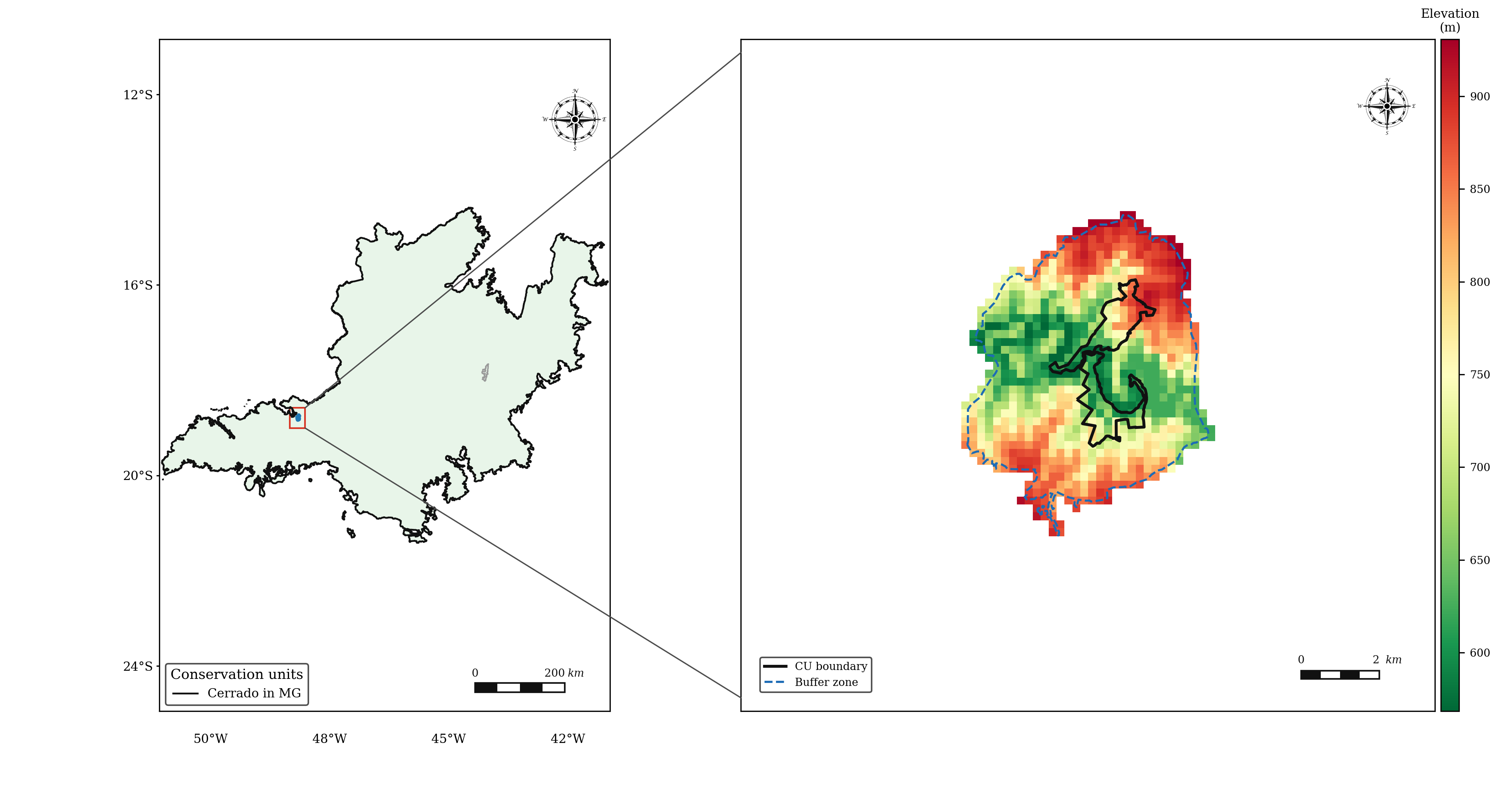}
    \textbf{(a)} Parque Estadual do Pau Furado
  \end{minipage}

  \vspace{0.35em}

  \begin{minipage}[t]{0.94\linewidth}
    \centering
    \includegraphics[width=\linewidth]{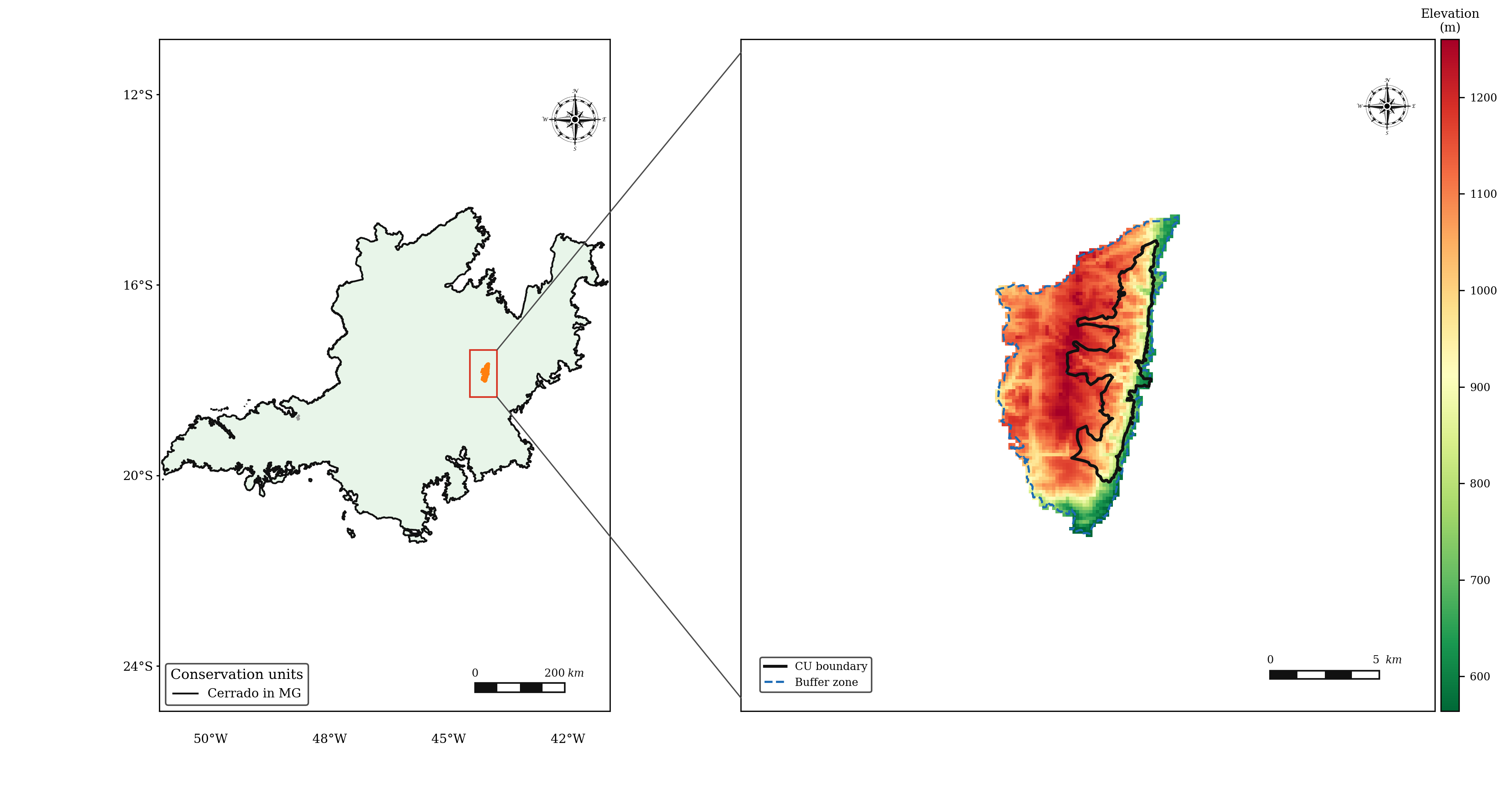}
    \textbf{(b)} Parque Estadual da Serra do Cabral
  \end{minipage}
  \caption{Held-out AOIs. Each panel locates the AOI inside Cerrado-MG and shows Copernicus DEM over the CU and official buffer zone.}
  \label{fig:aoi-locators}
\end{figure}

Two state CUs define the independent AOI test geography: Parque Estadual do Pau Furado, in Uberlândia, Minas Gerais, and Parque Estadual da Serra do Cabral, in Augusto de Lima, Minas Gerais.
Each CU is paired with its official buffer zone polygon distributed by the Infraestrutura de Dados Espaciais do Sistema Estadual de Meio Ambiente (IDE-Sisema).
The global training base excludes the union of these CU and buffer-zone polygons, while the AOI test set is sampled inside them.
Figure~\ref{fig:aoi-locators} locates each AOI within Cerrado-MG and zooms to the CU plus buffer-zone elevation surface used to describe local terrain context.

\FloatBarrier

\subsection{Fire labels and study period}

Positive labels come from the INPE Programa Queimadas BDQueimadas yearly comma separated files of reference-satellite active fire detections~\citep{inpebdq}.
The selected reference satellite is AQUA\_M-T, the early afternoon MODIS overpass on Aqua, whose 1 km contextual thermal-anomaly algorithm has recorded daily global active fire detections since 2002 and remains a widely used resource for fire research at regional and global scales~\citep{giglio2003enhanced,giglio2016collection6}.
Active fire products carry sensor-specific commission and omission behavior~\citep{libonati2015algorithm,rodrigues2019burned} so using one reference family avoids mixing distinct detection kernels.
The initial study period is 2019-01-01 to 2025-12-31, covering seven complete Cerrado dry seasons.

The raw archive was filtered to records whose biome and state fields identify Cerrado and Minas Gerais, then constrained to the study period and normalized to unique \texttt{(date, latitude, longitude)} tuples.
Each surviving point was spatially joined to the Cerrado-MG mask in a metric equal-area coordinate reference system so that all distance operations are computed in meters.
This first spatial screen removes records that match the tabular state and biome fields but fall outside the final polygonal footprint used by raster extraction.

\begin{figure}[!htbp]
  \centering
  \includegraphics[width=\linewidth]{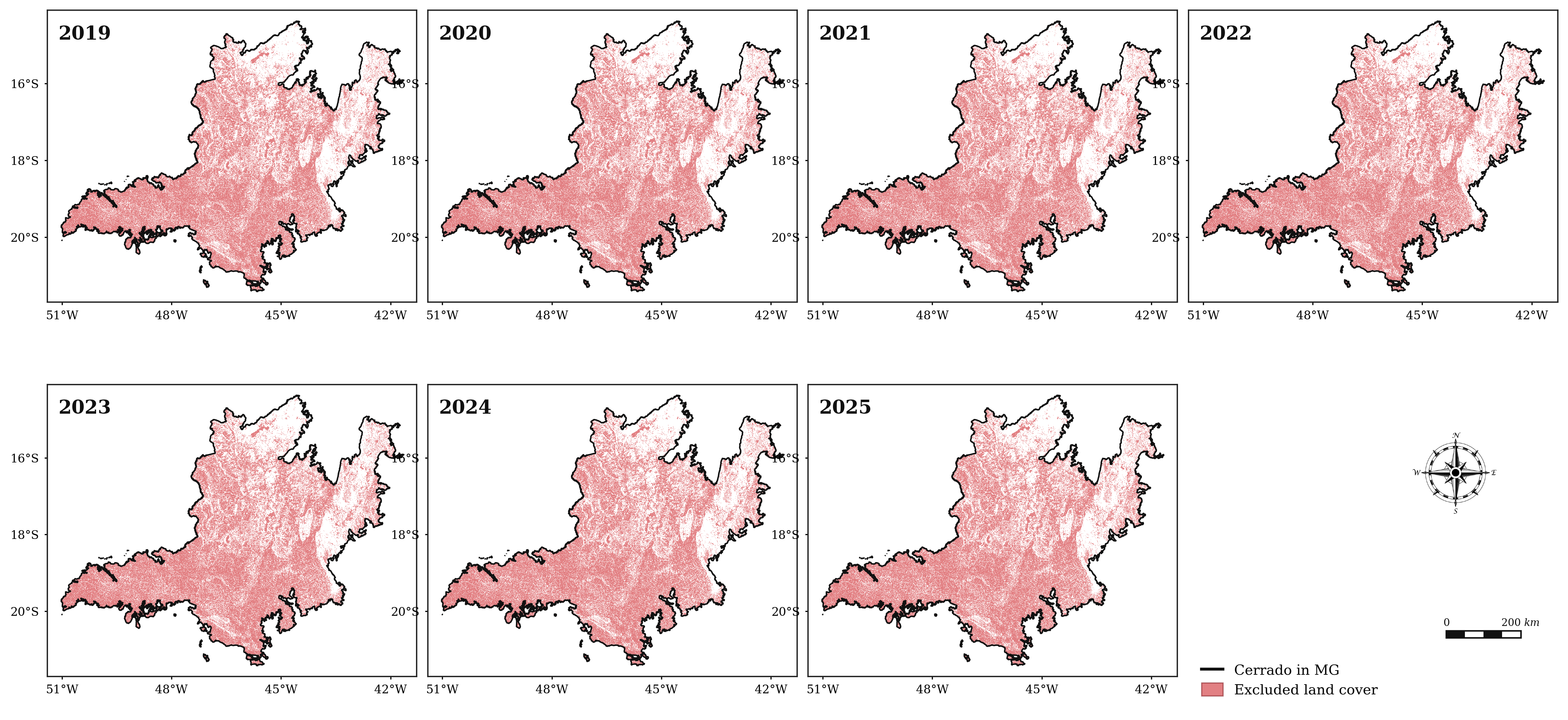}
  \caption{Dynamic MapBiomas Collection 9 exclusion mask by year. Excluded pixels are removed from same-year positives and pseudo-absence candidates.}
  \label{fig:eda-exclusion}
\end{figure}

\begin{figure}[!htbp]
  \centering
  \includegraphics[width=\linewidth]{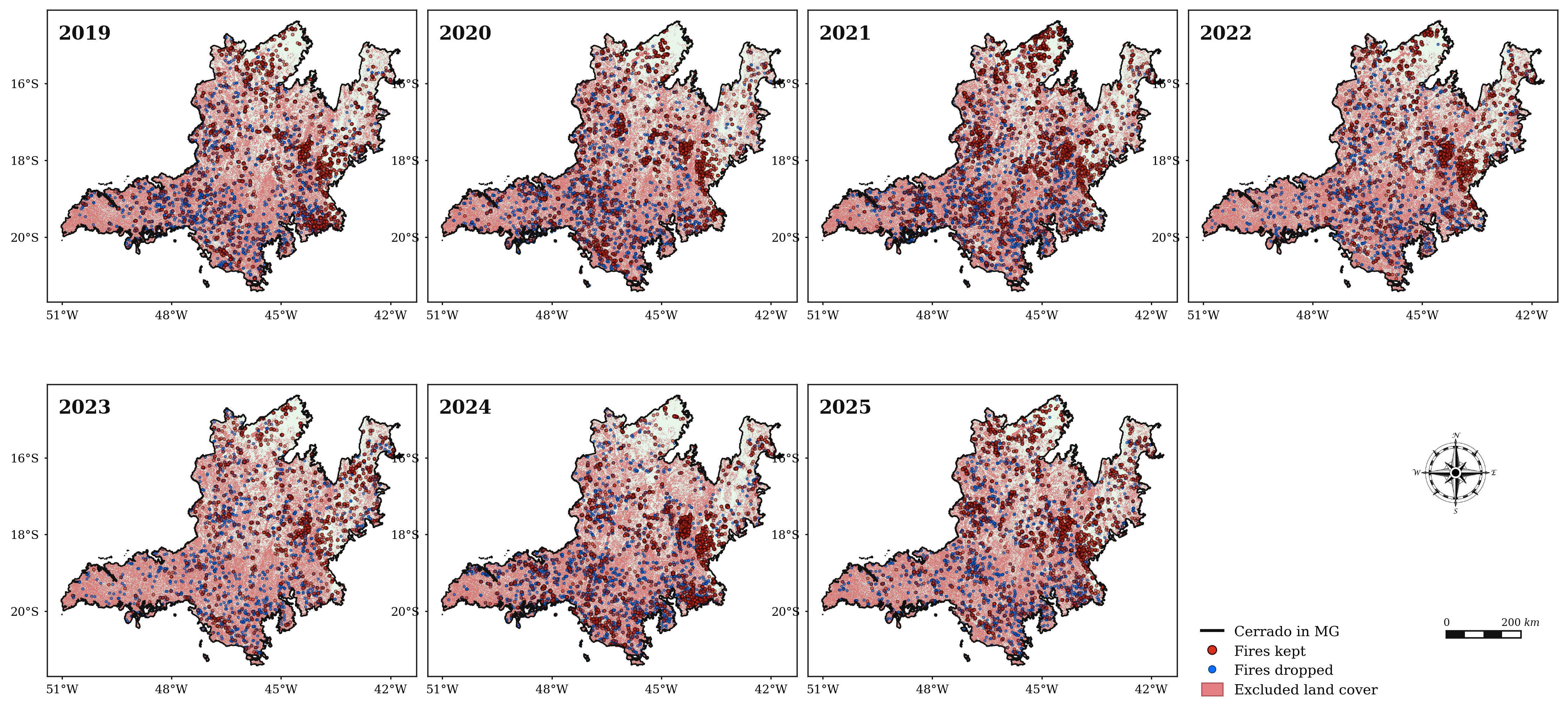}
  \caption{Positive-label screening after land-cover filtering. Red points are retained; blue points intersect excluded same-year MapBiomas classes.}
  \label{fig:eda-filtered}
\end{figure}

Land-cover filtering uses MapBiomas Collection 9 as an annual dynamic mask rather than a single static land-cover year.
For each fire point, the MapBiomas band corresponding to the fire year is sampled in Google Earth Engine, with the most recent available band used for later years when necessary~\citep{gorelick2017gee}.
Anthropogenic, mining, urban, aquaculture, and water classes are excluded before modeling.
The purpose is to keep the positive class aligned with natural Cerrado fire dynamics and to avoid temporal leakage: a 2019 fire is evaluated against the 2019 land-cover band, not against a later land-cover state.
Figure~\ref{fig:eda-exclusion} shows the excluded land-cover footprint by year, and Figure~\ref{fig:eda-filtered} shows the corresponding retained and dropped fire detections.

\FloatBarrier

\FloatBarrier

Multiple satellite detections can correspond to the same physical fire event because of overpass geometry, sub-pixel ambiguity, or repeated detection over consecutive days~\citep{hantson2013strengths,andela2019fireatlas}.
To reduce this redundancy, retained detections within three kilometers and three days of each other were merged and represented by the first detected event head, with both distances evaluated in the metric CRS~\citep{yang2026forests}.
The merged event heads retain their original date and coordinates, which preserves the daily binary classification target rather than aggregating the response into a long-term susceptibility surface.

\begin{figure}[!htbp]
  \centering
  \begin{minipage}[t]{0.55\linewidth}
    \centering
    \includegraphics[width=\linewidth]{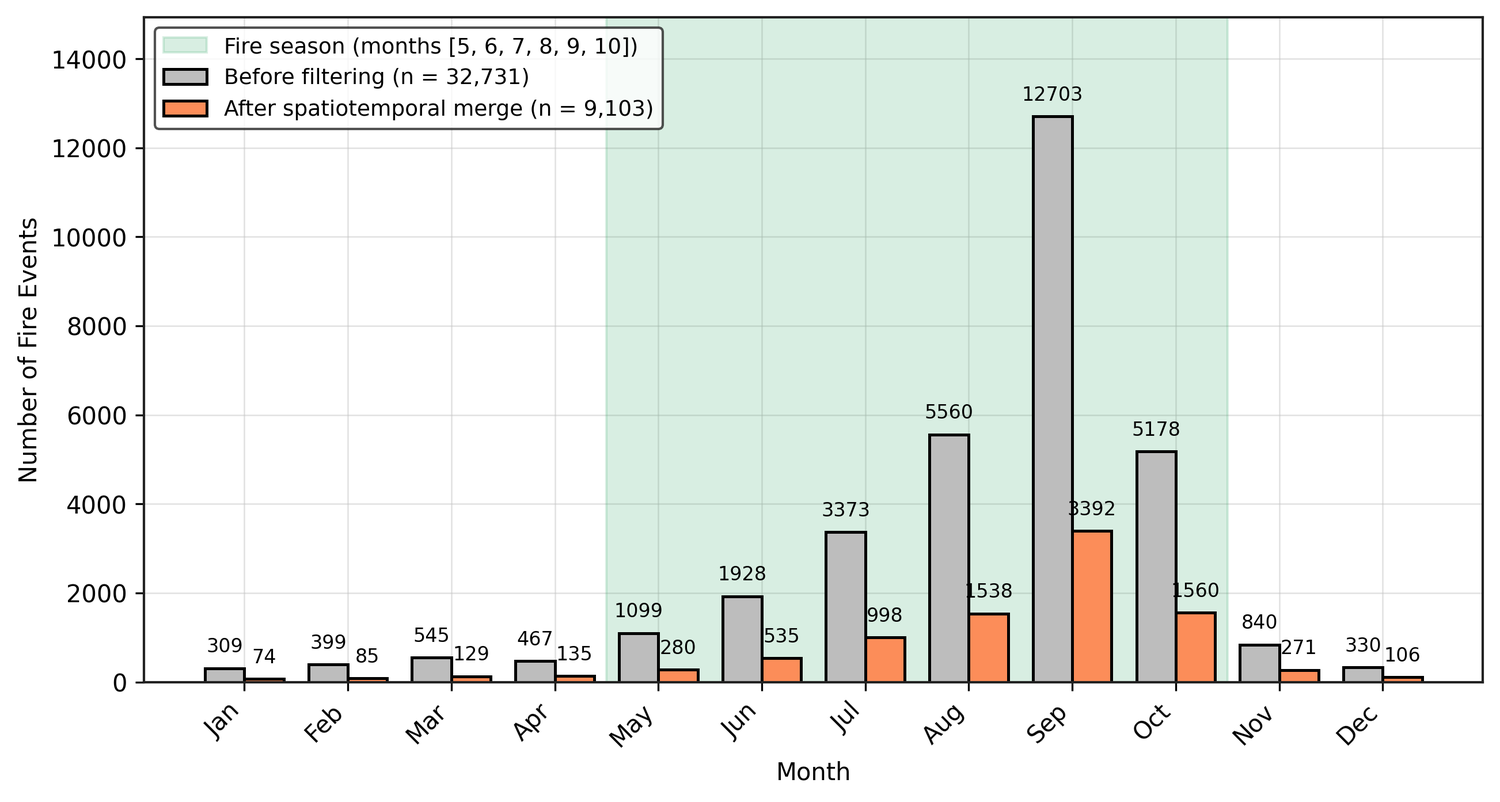}
  \end{minipage}\hfill
  \begin{minipage}[t]{0.39\linewidth}
    \centering
    \includegraphics[width=\linewidth]{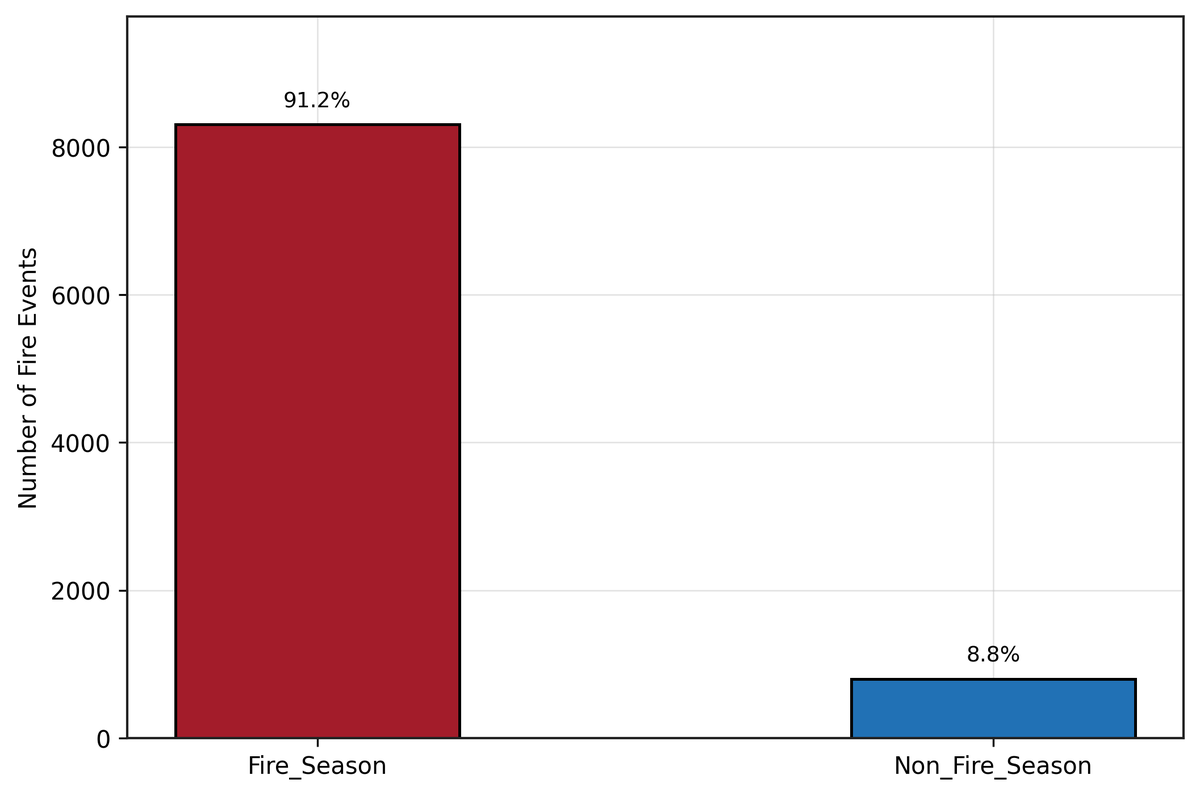}
  \end{minipage}
  \caption{Fire-season diagnostics for BDQueimadas detections inside Cerrado-MG. Panels show monthly counts and the May through October dry-season share.}
  \label{fig:fire-season-diagnostics}
\end{figure}

Finally, retained detections were restricted to May through October, the Cerrado dry-season window used for both positive labels and pseudo absences.
As shown in Figure~\ref{fig:fire-season-diagnostics}, the overwhelming majority of detections inside Cerrado-MG over the 2019 to 2025 study period occurred during these months, with a pronounced peak between July and September.
This concentration is consistent with regional climate seasonality and with the dry-spell driven burning propensity reported for the Cerrado biome~\citep{dieleviegas2022brazilianbiomes,hoffmann2020fire,setzer2019riscofogo}.

\FloatBarrier

\begin{figure}[!htbp]
  \centering
  \includegraphics[width=\linewidth]{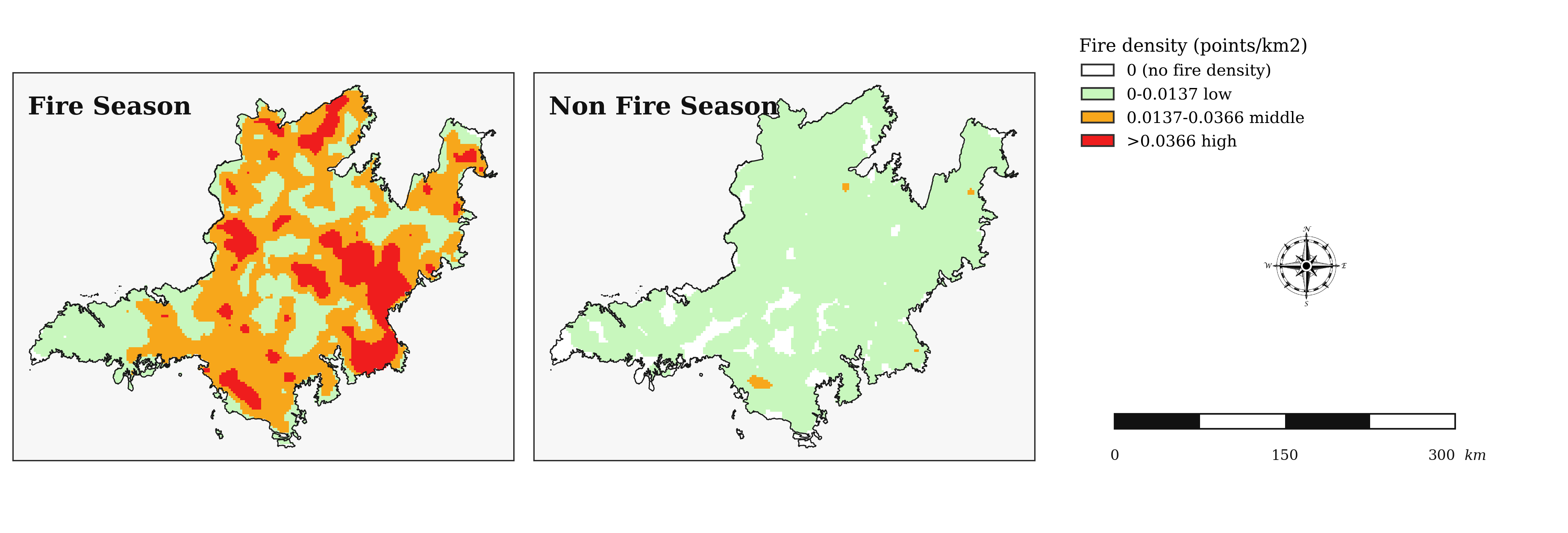}
  \caption{Fire-density contrast between the May through October fire season and the remaining months inside Cerrado-MG.}
  \label{fig:fire-density-season}
\end{figure}

The same seasonal restriction also has a spatial consequence.
Figure~\ref{fig:fire-density-season} shows that fire detections during the May through October window form the dominant density surface across Cerrado-MG, while detections outside that window are sparse by comparison.
Using the fire season for both positive labels and pseudo absences therefore keeps the classification problem focused on plausible fire-prone conditions instead of allowing easy background negatives from months with little observed burning.

\FloatBarrier

\subsection{Pseudo absences and AOI test samples}

Labels remain point-date events rather than coarse grid-cell aggregates, so class balance is handled at sampling time.
Training pseudo absences are sampled at ten background point-dates per positive and must satisfy five constraints: they lie inside the Cerrado-MG mask, fall on eligible MapBiomas land cover for their own calendar year, remain more than three kilometers from any same-day positive, occur outside the held-out CU and buffer-zone polygons, and fall inside the May through October fire-season window.
This contract keeps negatives physically comparable to positives and avoids easy background samples from irrelevant seasons or land covers~\citep{barbetmassin2012pseudo}.

The AOI test set is constructed independently from training.
Its positives are all retained BDQueimadas detections inside the two target CU and buffer-zone polygons over the study period.
Its pseudo absences are sampled inside the same AOI polygons under the same biome, land-cover, same-day distance, and season constraints, at one hundred absences per positive.
The same AOI point set is reused across all feature stages; only the covariate columns change.
This makes per-stage differences in AUC-PR interpretable as marginal feature value rather than as a consequence of different test samples.

\subsection{Covariate sources and staged extraction}

All covariates are extracted through Google Earth Engine, which supplies versioned image collections, reproducible server-side reducers, and spatially explicit sampling over the same point-date table~\citep{gorelick2017gee}.
The feature stages are nested: each stage contains the same rows as the previous stage plus additional covariates.
This row-alignment invariant lets the ablation compare atmospheric, surface, static spatial, and temporal memory families without changing labels, dates, coordinates, or AOI test identity.

The atmospheric baseline uses ERA5-Land daily aggregates for heat, humidity, wind, precipitation, and seasonal position.
Derived variables are defined in Table~\ref{tab:stage1-covariates} and use the transformations introduced in Section~\ref{sec:theory}.
This stage overlaps the meteorological logic of the INPE Risco de Fogo index while replacing hand tuned thresholds with learned classifiers.

\begin{table}[!htbp]
  \centering
  \small
  \caption{Stage 1 atmospheric covariates. All variables are sampled or derived for the target point-date from ERA5-Land daily aggregates before model fitting.}
  \label{tab:stage1-covariates}
  \begin{tabular}{p{0.22\linewidth}p{0.33\linewidth}p{0.35\linewidth}}
    \toprule
    Feature & Meaning & Source and derivation \\
    \midrule
    \texttt{t2m\_c} & Near-surface air temperature in degrees Celsius. & ERA5-Land \texttt{temperature\_2m}, daily mean, converted from kelvin to degrees Celsius. \\
    \texttt{rh2m} & Near-surface relative humidity in percent. & Derived from ERA5-Land two meter temperature and dewpoint temperature with the Tetens equation. \\
    \texttt{wind\_speed\_10m} & Ten meter wind speed in meters per second. & Derived from ERA5-Land zonal and meridional wind components. \\
    \texttt{wind\_dir\_sin}, \texttt{wind\_dir\_cos} & Circular representation of wind direction. & Derived from ERA5-Land wind components to avoid a discontinuity at 0 and 360 degrees. \\
    \texttt{precip\_mm} & Daily accumulated precipitation in millimeters. & ERA5-Land \texttt{total\_precipitation\_sum}, converted from meters to millimeters. \\
    \texttt{sin\_doy}, \texttt{cos\_doy} & Cyclic position of the day within the annual fire season. & Derived from the sample date as sine and cosine of day of year. \\
    \bottomrule
  \end{tabular}
\end{table}
\FloatBarrier

The surface-fusion stage adds MODIS vegetation, water-stress, and thermal surface-state variables.
Because valid MODIS observations vary by product, cloud cover, and orbit, extraction uses source-specific recent-valid compositing and records thermal observation age.
This is an extraction provenance strategy, not post hoc statistical imputation.

\begin{table}[!htbp]
  \centering
  \small
  \caption{Stage 2 surface covariates. Stage 2 is a strict superset of Stage 1 and adds independently composited MODIS optical and thermal variables.}
  \label{tab:stage2-covariates}
  \begin{tabular}{p{0.22\linewidth}p{0.33\linewidth}p{0.35\linewidth}}
    \toprule
    Feature & Meaning & Source and derivation \\
    \midrule
    \texttt{ndvi} & Vegetation greenness and photosynthetic activity. & MODIS Collection 6.1 MOD13Q1 \texttt{NDVI}, scaled by 0.0001 and sampled from the most recent valid pixel within the configured lookback window. \\
    \texttt{ndwi} & Vegetation and surface water-stress proxy. & Derived from MODIS Collection 6.1 MOD09GA near infrared and shortwave infrared reflectance as \((NIR - SWIR)/(NIR + SWIR)\). \\
    \texttt{lst} & Daytime land surface temperature in kelvin. & MODIS Collection 6.1 MOD11A1 \texttt{LST\_Day\_1km}, scaled by 0.02 and sampled from the most recent valid clear-sky pixel. \\
    \bottomrule
  \end{tabular}
\end{table}
\FloatBarrier

The static spatial stage adds terrain and human-access context that does not vary by date.
These variables test whether residual spatial structure remains after weather and surface state are already known, as suggested by multi-source geographic factor studies~\citep{yang2026forests,freitas2025triunfo}.

\begin{table}[!htbp]
  \centering
  \small
  \caption{Stage 3 static spatial covariates. Stage 3 is a strict superset of Stage 2 and adds terrain and built-environment context without changing the point-date table.}
  \label{tab:stage3-covariates}
  \begin{tabular}{p{0.22\linewidth}p{0.33\linewidth}p{0.35\linewidth}}
    \toprule
    Feature & Meaning & Source and derivation \\
    \midrule
    \texttt{slope} & Local terrain inclination in degrees. & Derived in Google Earth Engine from the SRTM \texttt{USGS/SRTMGL1\_003} elevation band and sampled at the model point. \\
    \texttt{aspect} & Local terrain orientation in degrees. & Derived in Google Earth Engine from the SRTM elevation band and sampled at the model point. \\
    \texttt{distance\_built\_m} & Distance to built-up land cover in meters. & Computed as capped Euclidean distance to ESA WorldCover 2021 class 50 pixels, using the configured 100 m distance-transform scale. \\
    \bottomrule
  \end{tabular}
\end{table}
\FloatBarrier

The temporal-memory stage adds causal EWMAs of selected atmospheric and surface variables at fixed half-lives of 3, 7, and 15 days.
The summaries use trailing history windows only, so they represent environmental memory available at the target date.
The grid is a short, parsimonious alternative to model-tuned lagged predictors and is motivated by antecedent rainfall logic, fire-danger systems with multiple fuel-moisture response classes, and applied fire or fuel-moisture models using lagged meteorological summaries~\citep{vanwagner1987fwi,dejong2016fwi,cavalcante2021firehazard,vilchisfrances2021daily,shmuel2022timelagged,hou2024fuelmoisture,setzer2019riscofogo}.

\begin{table}[!htbp]
  \centering
  \small
  \caption{Stage 4 temporal-memory covariates. Stage 4 is a strict superset of Stage 3 and adds causal exponentially weighted summaries of recent atmospheric and surface history.}
  \label{tab:stage4-covariates}
  \begin{tabular}{p{0.30\linewidth}p{0.30\linewidth}p{0.30\linewidth}}
    \toprule
    Feature family & Meaning & Source and derivation \\
    \midrule
    \texttt{t2m\_c\_ewma\_*d} & Recent thermal memory at half-lives of 3, 7, and 15 days. & Computed locally from trailing ERA5-Land daily temperature history over days \(D-1\) through \(D-60\). \\
    \texttt{rh2m\_ewma\_*d} & Recent humidity memory at half-lives of 3, 7, and 15 days. & Computed locally from trailing ERA5-Land derived relative humidity history. \\
    \texttt{precip\_mm\_ewma\_*d} & Recent precipitation memory at half-lives of 3, 7, and 15 days. & Computed locally from trailing ERA5-Land daily precipitation history. \\
    \texttt{ndvi\_ewma\_*d} & Recent vegetation-greenness memory at half-lives of 3, 7, and 15 days. & Computed locally from trailing MODIS NDVI history extracted with the same recent-valid compositing rule as Stage 2. \\
    \texttt{ndwi\_ewma\_*d} & Recent water-stress memory at half-lives of 3, 7, and 15 days. & Computed locally from trailing MODIS-derived NDWI history. \\
    \texttt{lst\_ewma\_*d} & Recent surface-temperature memory at half-lives of 3, 7, and 15 days. & Computed locally from trailing MODIS daytime LST history extracted from valid clear-sky pixels. \\
    \bottomrule
  \end{tabular}
\end{table}
\FloatBarrier

\section{Proposed method}
\label{sec:methods}
The protocol proceeds through row-aligned ablation tables, a three-model classifier suite, temporal validation, independent AOI transfer testing, and retrospective score mapping over dense daily grids.

\begin{figure}[!htbp]
  \centering
  \includegraphics[width=\linewidth]{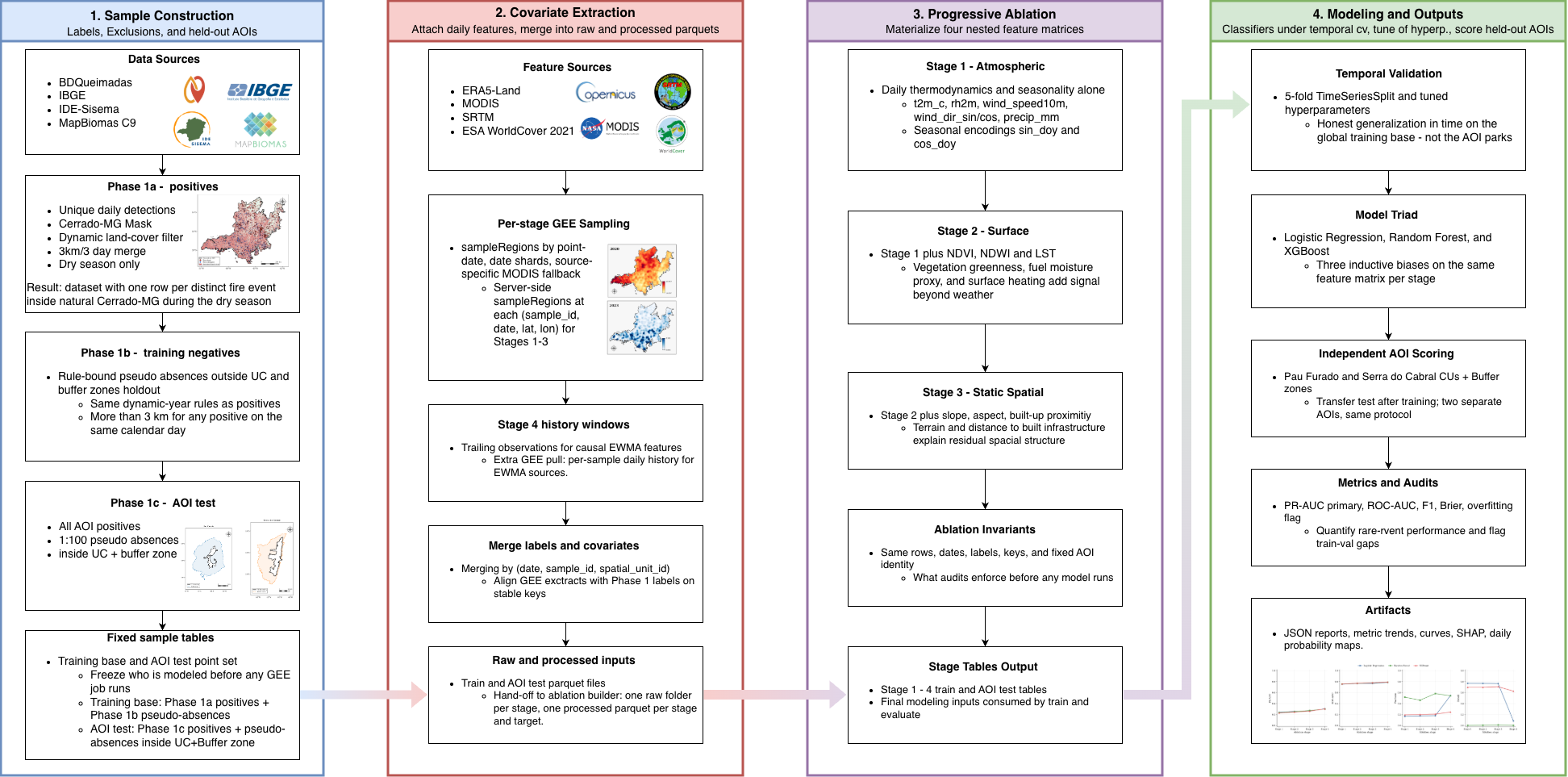}
  \caption{Technical workflow for the retrospective daily Cerrado active-fire detection evaluation.}
  \label{fig:technical-workflow}
\end{figure}

Figure~\ref{fig:technical-workflow} summarizes the complete technical workflow.

\FloatBarrier

\subsection{Progressive ablation design}

The ablation has four nested stages.
Let $D_k$ denote the modeling table for stage $k$, where each row is one labeled point-date tuple and each column after the identifiers is a predictor available to the classifier.
The ablation is progressive: $D_{k+1}$ must preserve all rows, dates, labels, and identifiers in $D_k$, and its feature set must be a strict superset of the previous stage.
Consequently, a performance change between two stages is attributable to the added variable family rather than to a new sample, a different time window, or a different prevalence.
The same invariant is enforced for the AOI test tables, where the point set is fixed once and reused from Stage 1 through Stage 4.

Stage 1 is the atmospheric baseline and includes daily temperature, relative humidity, wind, precipitation, and seasonal sine and cosine terms derived from day of year.
It tests whether daily thermodynamic and rainfall conditions alone are sufficient to rank fire-prone point-dates.
Stage 2 adds optical and thermal surface state through NDVI, NDWI, and land surface temperature, testing whether vegetation greenness, fuel water stress, and near-surface heating reduce errors left by the atmospheric baseline.
Stage 3 adds static spatial context through terrain and built-environment proximity, testing whether slope, aspect, and distance to built-up areas explain residual spatial structure after daily weather and surface state are known.
Stage 4 adds causal exponentially weighted moving averages of selected atmospheric and surface variables, testing whether short environmental memory improves daily ranking beyond instantaneous same-day conditions.

All stages are audited before modeling.
The audit checks row counts, key identity, label identity, column nesting, missingness, and AOI test stability.
Because the feature tables include same-day environmental covariates, the study is interpreted as retrospective daily classification and ranking rather than prospective forecasting.
Rows are not silently imputed or dropped to make a later stage easier to fit.
When a sensor-specific extraction gap remains unresolved, the corresponding stage table must expose the gap through the audit rather than hiding it inside the model matrix.

\subsection{Model suite and hyperparameter search}

For every stage, the same three classifiers are trained on the same feature matrix and evaluated under the same splits.
Logistic Regression is used as the linear reference model.
It tests whether each covariate stage contains an approximately monotonic and additive signal on the log-odds scale.
Random Forest is used as the bagging baseline for nonlinear thresholds and interactions.
XGBoost is used as the boosting baseline for higher-order interactions among climate, vegetation, terrain, and proximity variables~\citep{chen2016xgboost,yang2026forests,bian2024forests,freitas2025triunfo}.
The three models are compared under the same rows, features, folds, and AOI tests.

Hyperparameters are selected separately for each model and stage using Optuna with a Tree-structured Parzen Estimator sampler and a median pruner.
The tuning objective is the mean AUC-PR across the same five time-series folds later used for validation, so the selected hyperparameters are optimized for rare-event ranking rather than for accuracy or AUC-ROC.
The trial budget is fixed in the project configuration: 30 trials for Logistic Regression, 30 trials for Random Forest, and 50 trials for XGBoost, each with a per-trial timeout of 600 seconds.
Trials are run sequentially, while model-internal fitting uses the available processor threads.
After tuning, the best hyperparameters are refit on the complete global training table for the corresponding stage, and the fitted model is stored with its evaluation and diagnostic artifacts.

Class imbalance is handled through model-side weighting where supported and through evaluation metrics that remain meaningful under rare prevalence.
The training table uses the 1:10 positive to pseudo-absence contract described in Section~\ref{sec:data}, while XGBoost receives a stage-specific positive-class weight derived from the training labels.
The independent AOI test uses a more severe 1:100 positive to pseudo-absence design.
No synthetic over-sampling is used, because synthetic balancing would change the rare-event prevalence that the study is designed to expose.

\subsection{Temporal validation and independent AOI testing}

Model validation has two levels.
First, each stage and model pair is evaluated with five-fold time-series cross-validation on the global training base.
The split is ordered and unshuffled: each validation fold occurs later in time than its corresponding training fold.
This design prevents nearby dates from the same fire season from appearing simultaneously in training and validation sets, a leakage risk that is especially acute for daily environmental covariates and short-lived fire events.
For each fold, the pipeline records both training and validation metrics.

Second, the tuned model is refit on the complete global training base and scored on the independent AOI test set.
The two test geographies are Parque Estadual do Pau Furado and Parque Estadual da Serra do Cabral, each evaluated with its official buffer zone.
These CU and buffer-zone polygons are excluded from the global training base before pseudo absences are drawn, and the AOI test set is built separately inside the excluded polygons.
The AOI protocol measures spatial transfer into protected-area management surfaces.
Each AOI receives its own metric block because sample size and fire prevalence differ sharply between the two sites.

The AOI test is deliberately more imbalanced than the training table.
All retained BDQueimadas positives inside the AOI polygons are kept, and pseudo absences are sampled at one hundred background point-dates per positive under the same season, land-cover, and same-day distance rules.
This produces a test setting closer to a manager-facing alert surface, where most candidate point-days are non-fire days.
The same AOI point set is reused for all four stages, so Stage 2, Stage 3, or Stage 4 improvements cannot be explained by new test negatives or by changed positive locations.

\subsection{Metrics and diagnostic interpretation}

AUC-PR is the primary metric for both hyperparameter search and reporting.
Under rare daily active-fire detection occurrence, AUC-PR evaluates whether the model ranks observed active-fire detections above the large background of plausible non-fire point-dates, and it is more sensitive than AUC-ROC to the false-positive burden emphasized in fire occurrence model evaluation guidelines~\citep{davis2006prroc,saito2015precision,phelps2021guidelines}.
AUC-ROC is retained as a secondary discrimination metric, but it is not used as the headline criterion.
The present manuscript reports precision and recall at probability-score cutoff 0.5 to expose warning burden and missed-fire behavior without treating the cutoff as an optimized operational threshold.

Cross-validation reporting uses the five-fold validation means produced by the current evaluation artifacts.
Training metrics are retained in the JSON reports as fit diagnostics, but the manuscript emphasizes validation and AOI results rather than in-sample performance.
For the AOI test, each report stores AUC-PR, AUC-ROC, threshold precision and recall, class counts, and a low-positive warning.
Class counts are interpreted together with AUC-PR, especially for low-support AOIs.

\subsection{Interpretability and spatial outputs}

Interpretability is produced after fitting and is not used for model selection.
Each trained model emits a feature-importance artifact, and SHAP values are computed on the AOI test rows, with subsampling when the test table exceeds the configured limit.
The objective is to inspect model behavior in relation to physically interpretable drivers such as heat, humidity, recent rainfall, vegetation condition, water stress, terrain, and proximity to built-up areas.
Feature-importance and SHAP summaries are standard interpretability tools in recent wildfire studies~\citep{cilli2022xai,yang2026forests,freitas2025triunfo}.
In the final analysis, they are read together with fold metrics, AOI metrics, and local false-positive or false-negative examples.

The paper also distinguishes point-test performance from mapped behavior.
Point tests evaluate ranking on sampled AOI positives and pseudo absences, while maps show where a fitted model concentrates high scores over a continuous protected-area surface.
Spatial score maps and temporal mirror plots are diagnostic artifacts used to inspect spatial plausibility, warning-volume concentration, and score changes near observed fire dates.

\subsection{Operational-style retrospective simulation}

The final experiment simulates a daily decision-support surface over the two held-out AOIs.
For each selected fire-season window, a dense grid is generated across the CU and buffer-zone polygons.
The grid spacing is 500 m by default, and every grid cell is paired with every date in the selected window.
The same Stage 1 through Stage 4 covariate extraction and preprocessing logic used for the ablation tables is then applied to these grid-date rows.
The trained Stage 4 Logistic Regression, Random Forest, and XGBoost models score each grid-date independently.

This simulation is retrospective, not a prospective forecast, because the current baseline uses same-day covariates unless a lagged-feature configuration is explicitly selected.
Stage 4 temporal memory remains causal because EWMA features are computed from previous history windows rather than future observations.
The experiment tests whether trained models concentrate elevated scores near observed fire activity without assigning uniformly high warnings across the full CU plus buffer-zone surface.

\section{Results}
\label{sec:results}

Results are reported in five parts: temporal cross-validation, Stage 4 explanatory profiles, independent AOI test performance, AOI-specific explanations, and retrospective operational-style maps.

\subsection{Cross-validation performance across ablation stages}

\begin{figure}[!htbp]
  \centering
  \includegraphics[width=\linewidth]{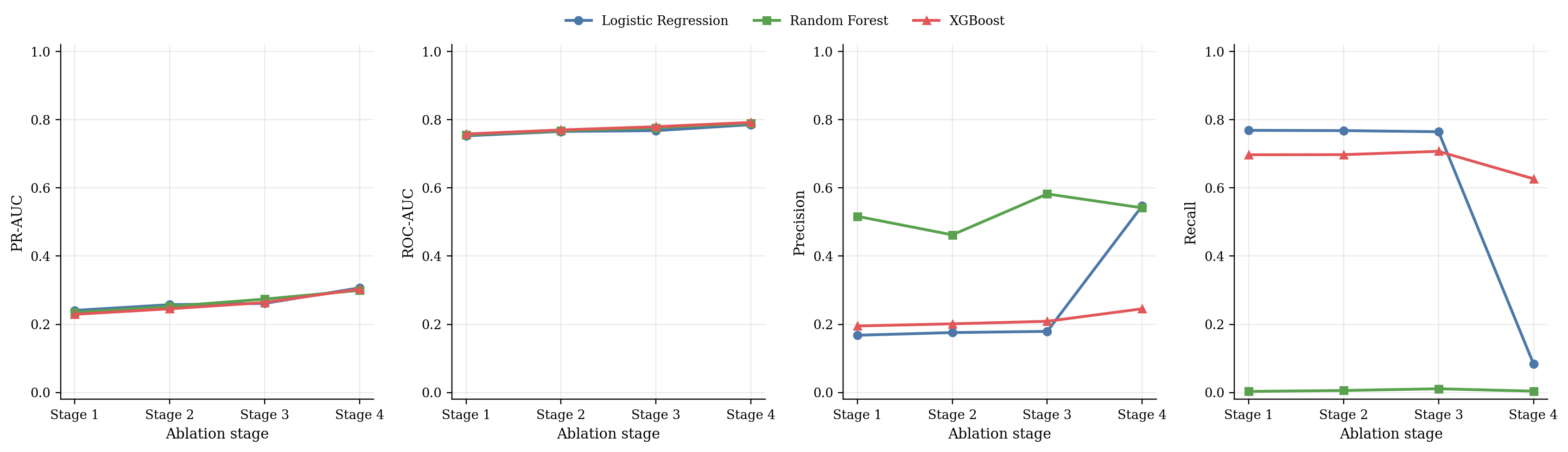}
  \caption{\footnotesize Temporal-validation metrics by ablation stage and model.}
  \label{fig:cv-metric-trends}
\end{figure}

Temporal validation shows a monotonic increase in mean AUC-PR from the atmospheric baseline to the temporal-memory stage for all three model families (Figure~\ref{fig:cv-metric-trends}).
In Stage 1, mean validation AUC-PR was 0.241 for Logistic Regression, 0.235 for Random Forest, and 0.229 for XGBoost.
Stage 2 increased these values to 0.258, 0.252, and 0.246, respectively, while Stage 3 reached 0.261, 0.274, and 0.264.
The complete Stage 4 temporal-memory table produced the highest cross-validation AUC-PR for each model, with values of 0.307 for Logistic Regression, 0.300 for Random Forest, and 0.303 for XGBoost.
The corresponding AUC-ROC values in Stage 4 were 0.786, 0.790, and 0.792, respectively.
Because the global training table uses a 1:10 positive to pseudo-absence design, the sampled no-skill AUC-PR baseline is approximately 0.091.
Stage 4 therefore improves ranking in the Cerrado-MG training domain; AOI transfer is evaluated separately.

\FloatBarrier

\begin{table}[!htbp]
  \centering
  \footnotesize
  \caption{Mean five-fold temporal cross-validation AUC-PR and AUC-ROC by ablation stage. Values summarize the fold-level validation metrics shown in Figure~\ref{fig:cv-metric-trends}.}
  \label{tab:cv-val-stage-metrics}
  \begin{tabular*}{\linewidth}{@{\extracolsep{\fill}}lcccccc@{}}
    \toprule
    Stage & LR AUC-PR & LR AUC-ROC & RF AUC-PR & RF AUC-ROC & XGB AUC-PR & XGB AUC-ROC \\
    \midrule
    1 & 0.241 & 0.753 & 0.235 & 0.756 & 0.229 & 0.758 \\
    2 & 0.258 & 0.766 & 0.252 & 0.768 & 0.246 & 0.770 \\
    3 & 0.261 & 0.768 & 0.274 & 0.776 & 0.264 & 0.779 \\
    4 & 0.307 & 0.786 & 0.300 & 0.790 & 0.303 & 0.792 \\
    \bottomrule
  \end{tabular*}
\end{table}
\FloatBarrier

\begin{figure}[!htbp]
  \centering
  \includegraphics[width=0.78\linewidth]{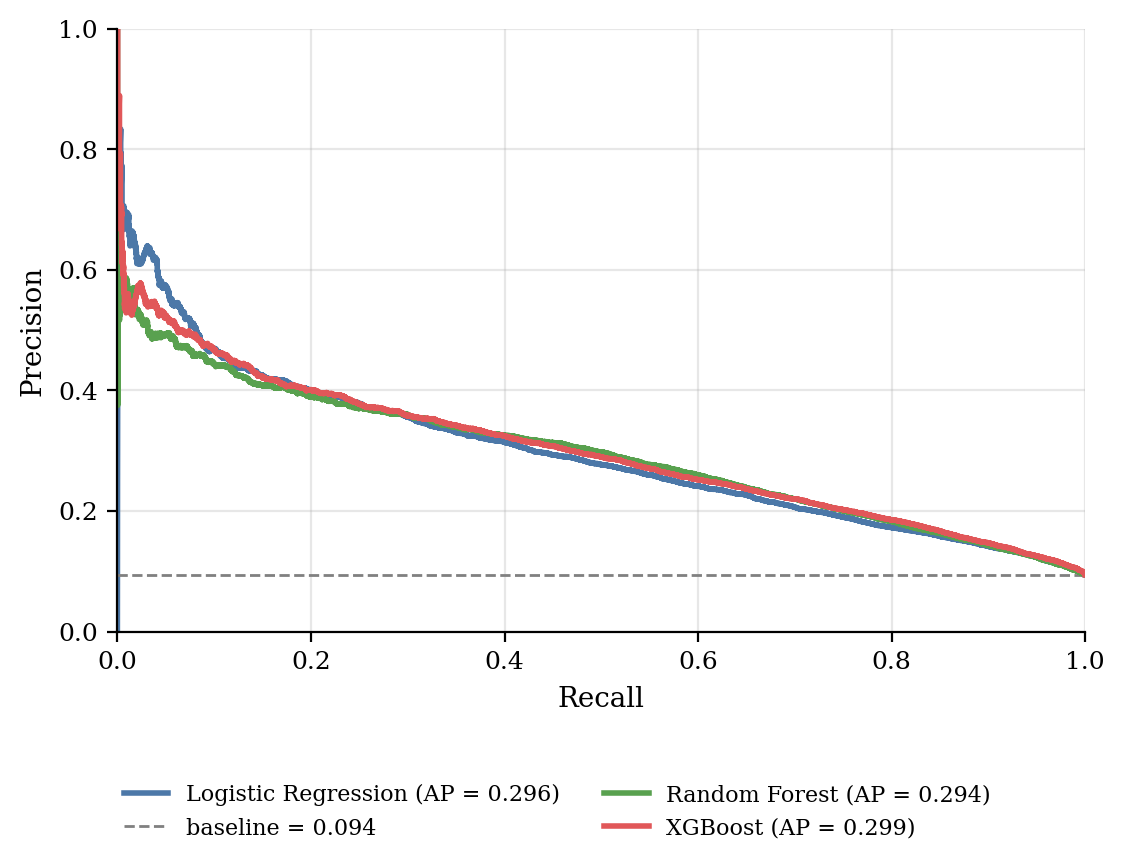}
  \caption{\footnotesize Stage 4 precision-recall curves on pooled temporal-validation predictions.}
  \label{fig:stage4-cv-curves-pr}
\end{figure}

\begin{figure}[!htbp]
  \centering
  \includegraphics[width=0.78\linewidth]{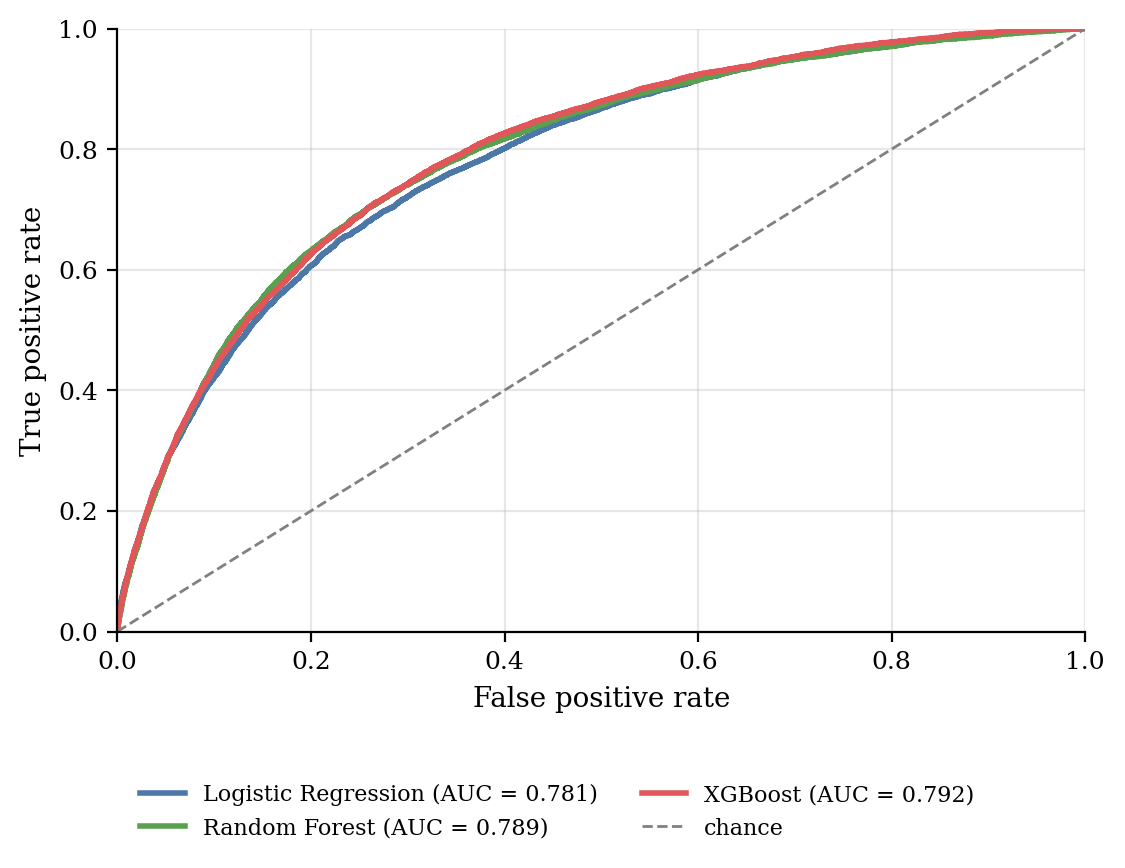}
  \caption{\footnotesize Stage 4 ROC curves on pooled temporal-validation predictions.}
  \label{fig:stage4-cv-curves-roc}
\end{figure}

Figure~\ref{fig:stage4-cv-curves-pr} gives the Stage 4 precision-recall traces on pooled temporal-validation predictions.
They separate the three model families more clearly than the corresponding ROC traces in Figure~\ref{fig:stage4-cv-curves-roc}, consistent with the metric hierarchy defined in Section~\ref{sec:methods}.

\FloatBarrier

\FloatBarrier

\begin{figure}[!htbp]
  \centering
  \begin{minipage}[t]{0.32\linewidth}
    \centering
    \includegraphics[width=\linewidth]{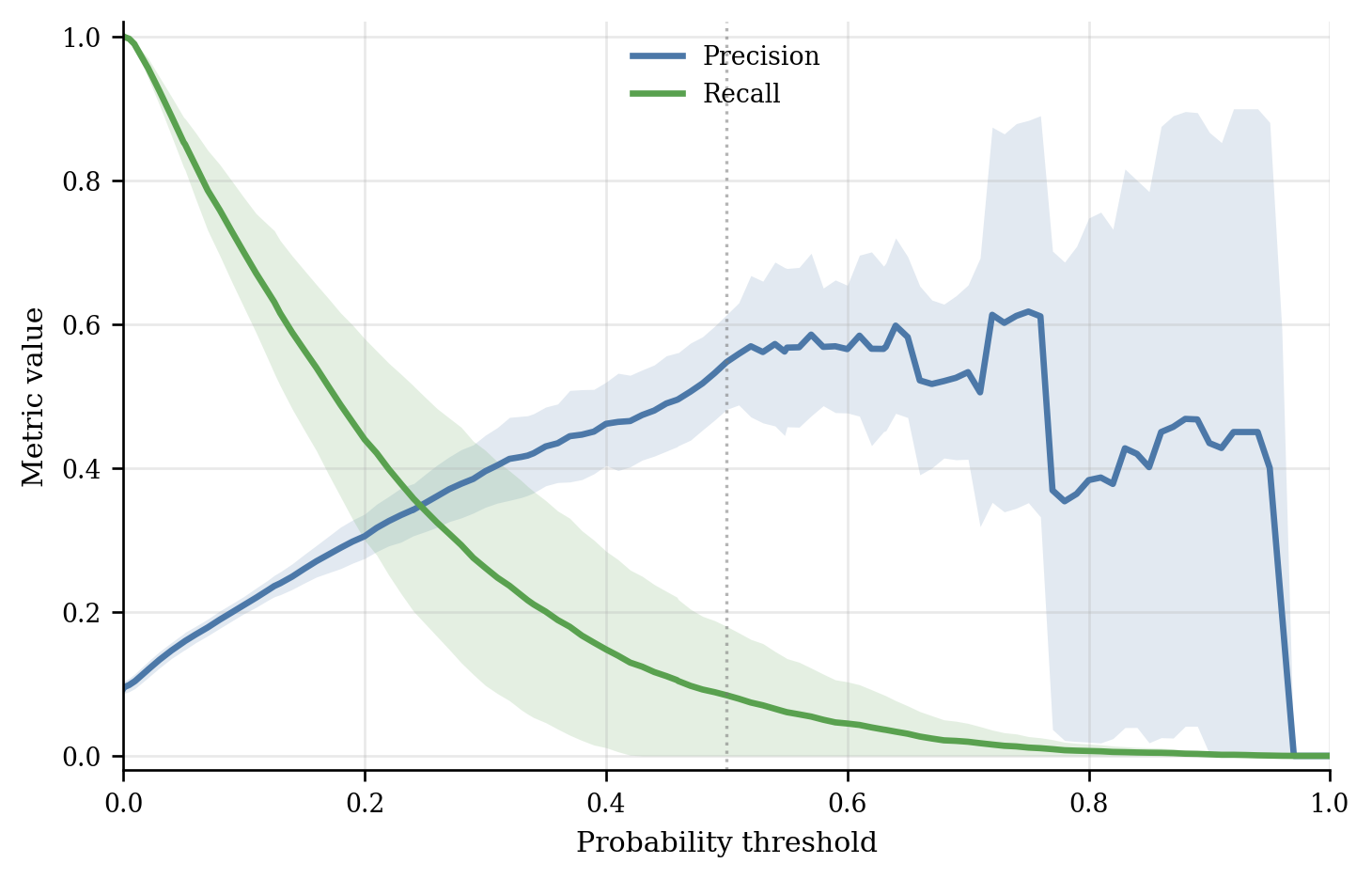}
    \scriptsize \textbf{(a)} Logistic Regression
  \end{minipage}\hfill
  \begin{minipage}[t]{0.32\linewidth}
    \centering
    \includegraphics[width=\linewidth]{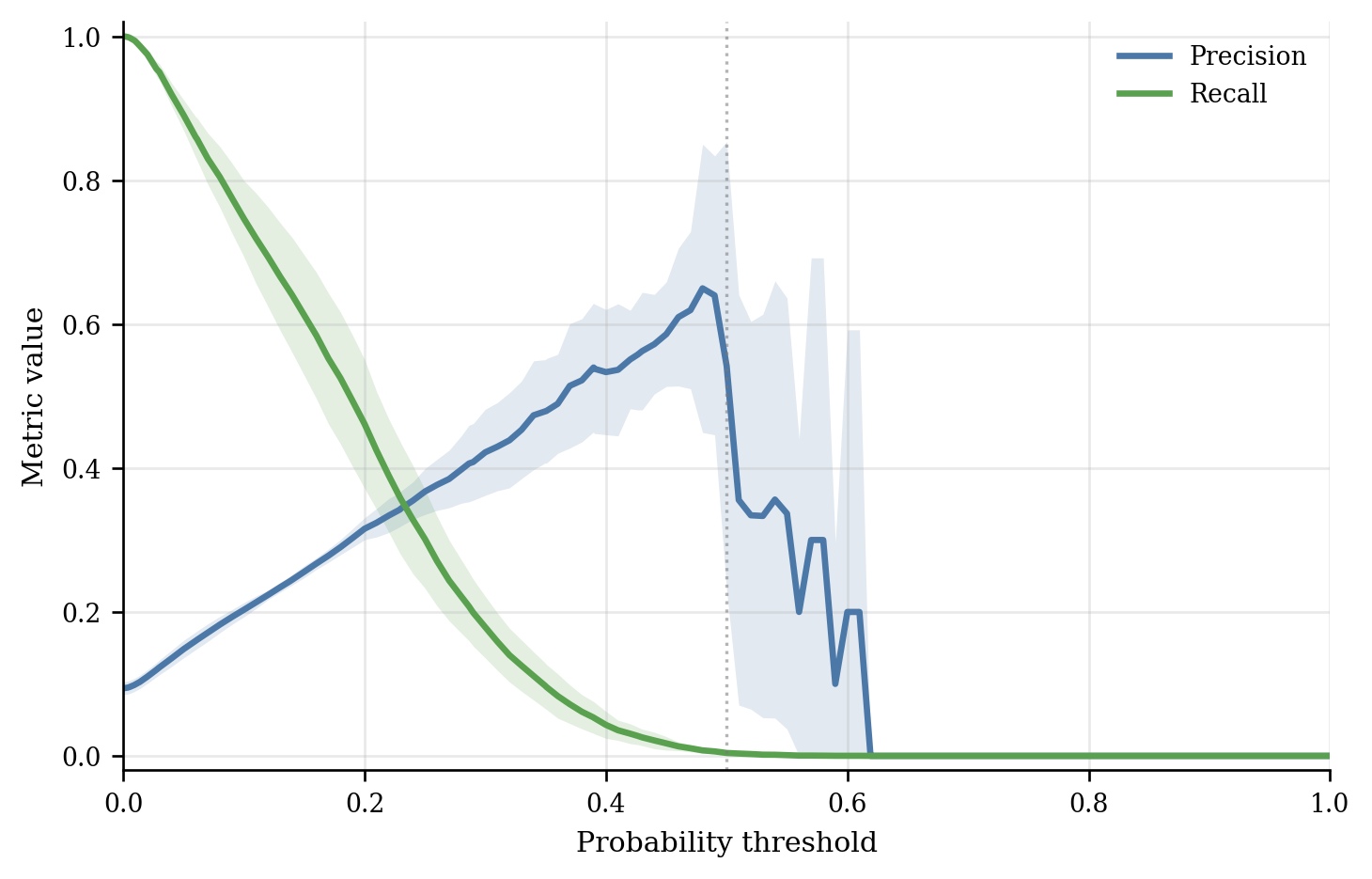}
    \scriptsize \textbf{(b)} Random Forest
  \end{minipage}\hfill
  \begin{minipage}[t]{0.32\linewidth}
    \centering
    \includegraphics[width=\linewidth]{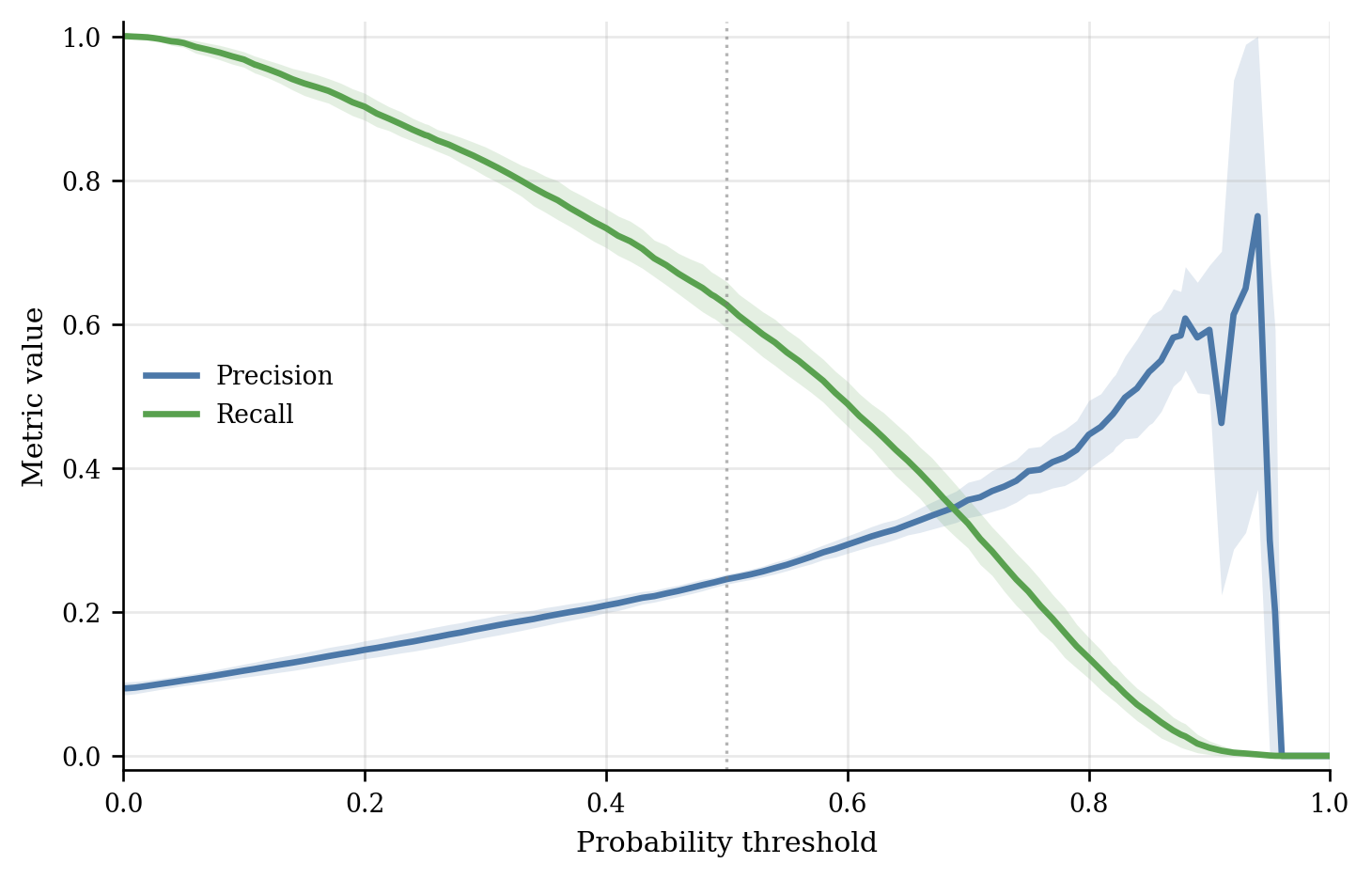}
    \scriptsize \textbf{(c)} XGBoost
  \end{minipage}
  \caption{\footnotesize Stage 4 temporal-validation Precision and Recall across probability-score thresholds. Shaded bands show 95 percent confidence intervals across folds.}
  \label{fig:stage4-cv-threshold-sensitivity}
\end{figure}

Figure~\ref{fig:stage4-cv-threshold-sensitivity} shows the corresponding Stage 4 threshold sensitivity for Precision and Recall.
Each panel summarizes the five temporal-validation folds with the fold mean and a 95 percent confidence interval.
The curves show the Precision-Recall tradeoff beyond the fixed 0.5 diagnostic cutoff.

\FloatBarrier

\subsection{Stage 4 explanatory profiles under temporal validation}

\begin{figure}[!htbp]
  \centering
  \begin{minipage}[t]{0.45\linewidth}
    \centering
    \includegraphics[width=\linewidth]{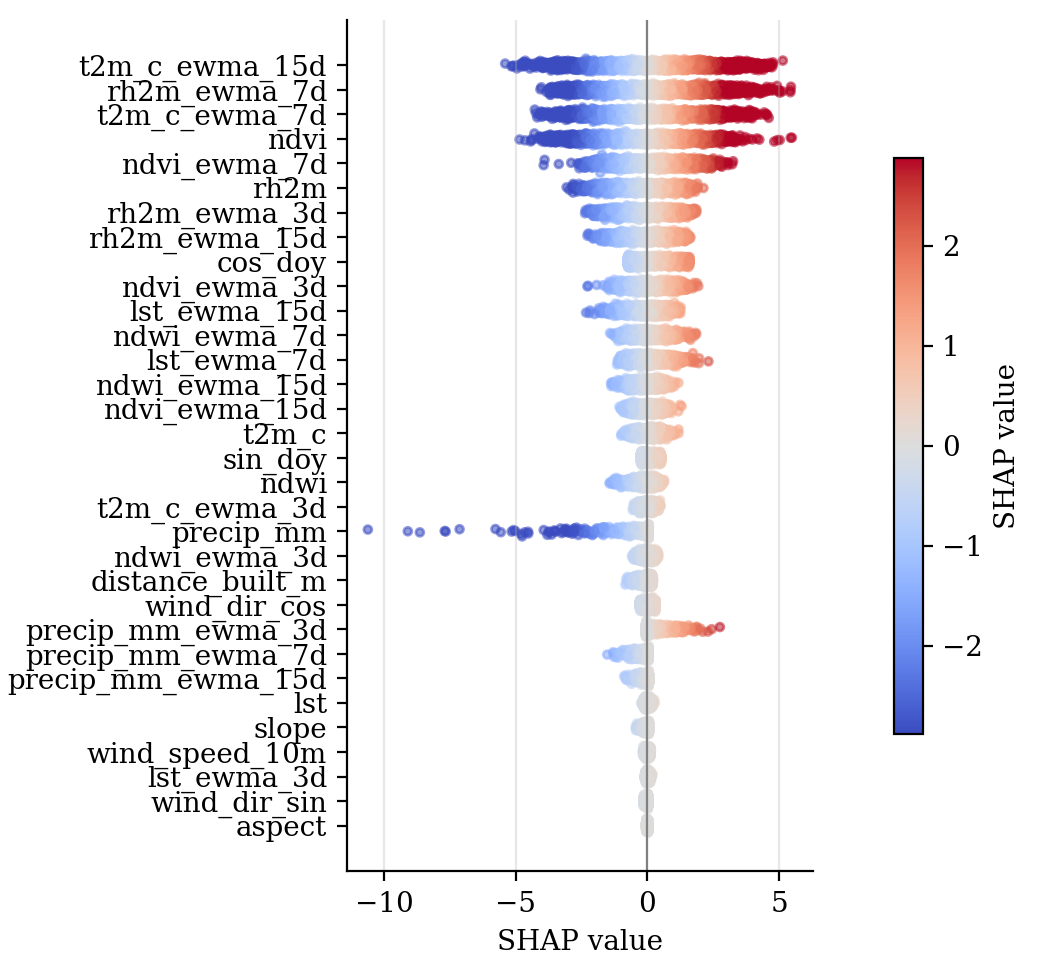}
    \scriptsize \textbf{(a)} Logistic Regression
  \end{minipage}\hfill
  \begin{minipage}[t]{0.45\linewidth}
    \centering
    \includegraphics[width=\linewidth]{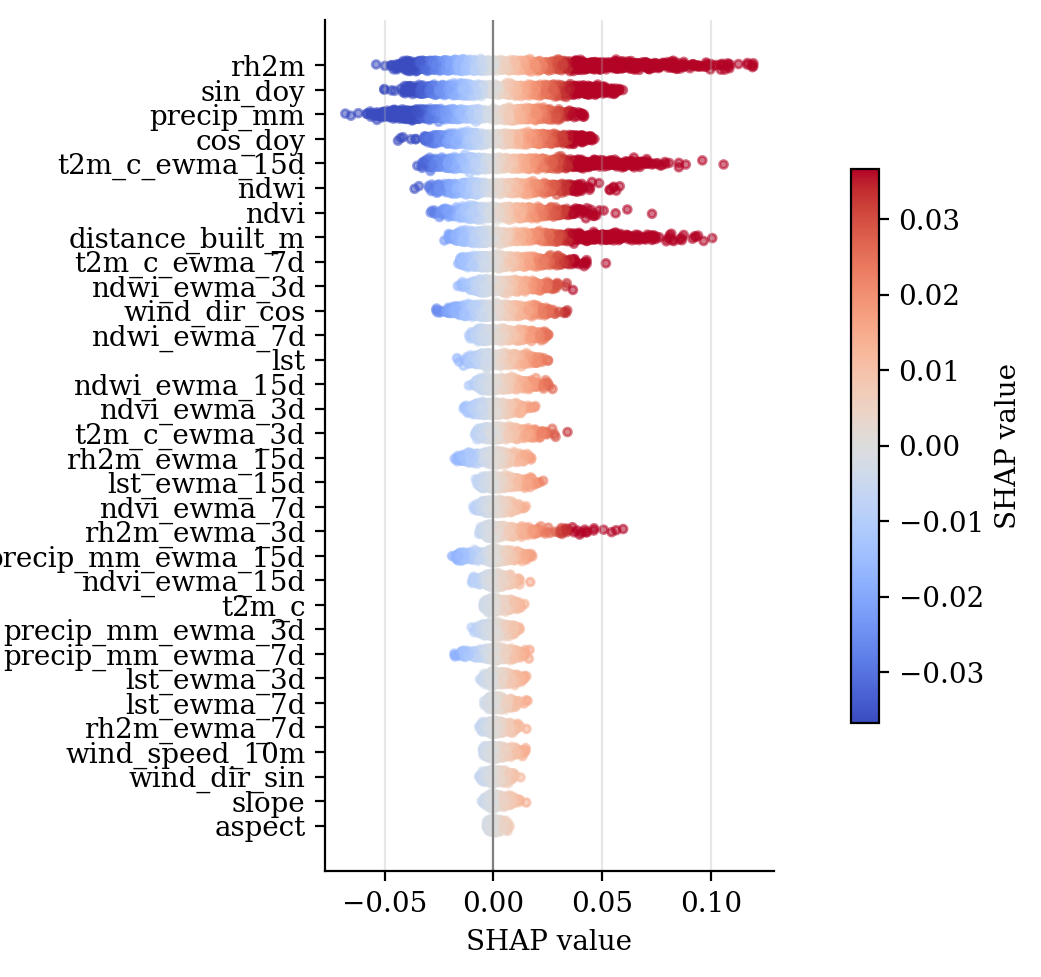}
    \scriptsize \textbf{(b)} Random Forest
  \end{minipage}
  \vspace{0.2em}

  \begin{minipage}[t]{0.45\linewidth}
    \centering
    \includegraphics[width=\linewidth]{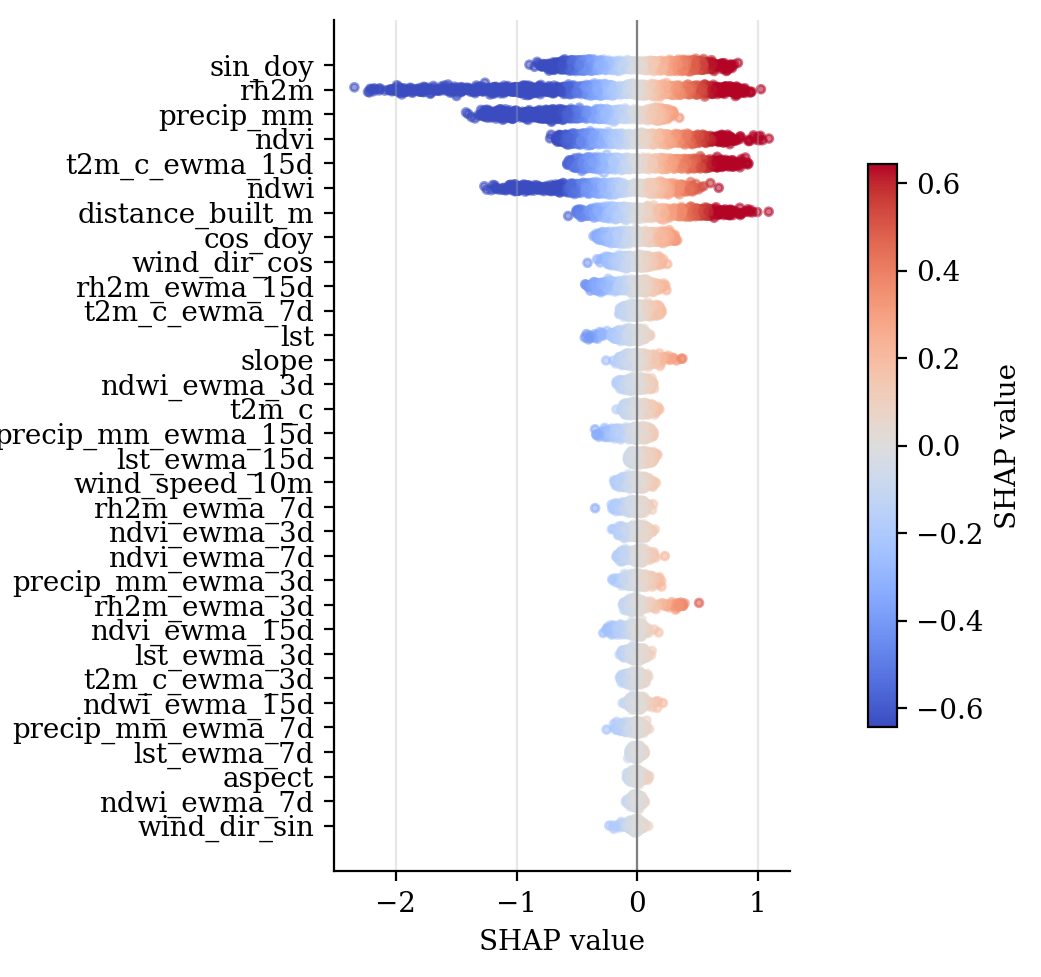}
    \scriptsize \textbf{(c)} XGBoost
  \end{minipage}
  \caption{\footnotesize Stage 4 temporal-validation SHAP summaries for (\textbf{a}) Logistic Regression, (\textbf{b}) Random Forest, and (\textbf{c}) XGBoost.}
  \label{fig:stage4-cv-shap}
\end{figure}

Stage 4 SHAP summaries were generated for the three trained model families under the cross-validation validation surface (Figure~\ref{fig:stage4-cv-shap}).
They show which Stage 4 covariates contributed most to fitted probability scores, separating instantaneous atmospheric and surface variables from the EWMA memory features.

\FloatBarrier

\subsection{Independent AOI test performance}

\begin{figure}[!htbp]
  \centering
  \includegraphics[width=\linewidth]{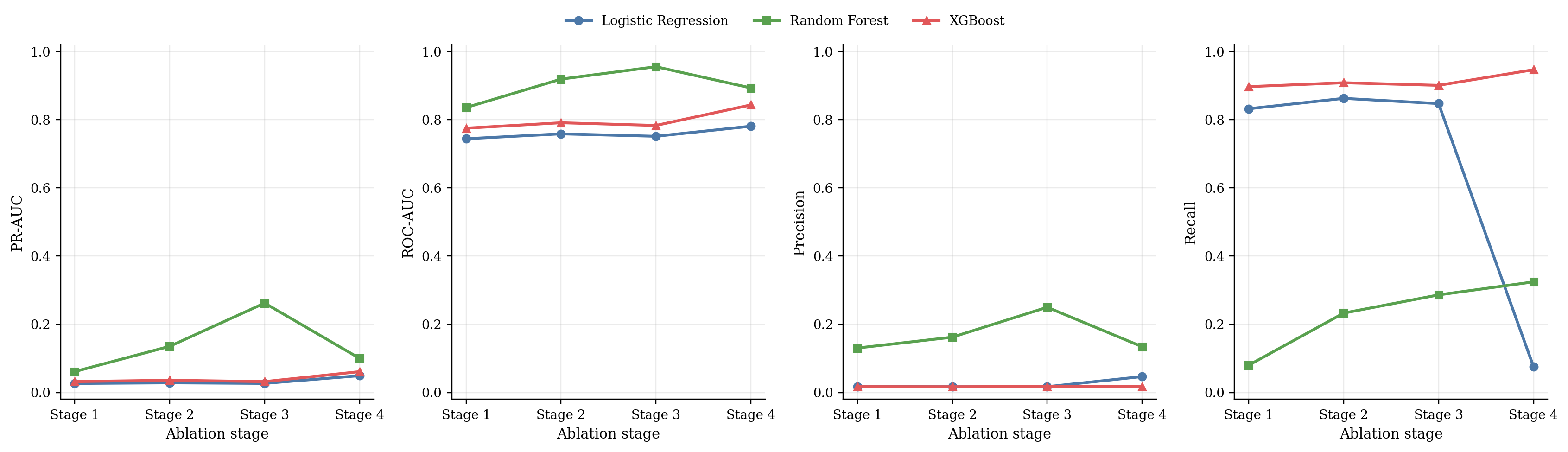}
  \caption{\footnotesize Serra do Cabral AOI metrics by stage and model (262 positives, 26,200 pseudo absences).}
  \label{fig:serra-cabral-aoi-trends}
\end{figure}

The independent AOI tests evaluate spatial transfer under a stricter 1:100 positive to pseudo-absence sampling ratio than the global training folds.
This design fixes the AOI no-skill AUC-PR baseline near 0.010.
Serra do Cabral is the primary AOI transfer reading because it contained 262 positives and 26,200 pseudo absences, providing a more stable held-out surface under the same prevalence design.
Random Forest reached the strongest Serra do Cabral AUC-PR in Stages 2 and 3, with values of 0.135 and 0.262, respectively, and remained the strongest Stage 4 model with AUC-PR of 0.100 (Figure~\ref{fig:serra-cabral-aoi-trends}).
At Stage 4, XGBoost obtained AUC-PR of 0.061 and Logistic Regression obtained AUC-PR of 0.049.
The Stage 4 threshold diagnostics differed across model families: Random Forest combined precision of 0.134 with recall of 0.324, while XGBoost produced recall of 0.947 with precision of 0.018.

\FloatBarrier

\begin{table}[!htbp]
  \centering
  \scriptsize
  \caption{Serra do Cabral AOI test metrics by ablation stage and model.}
  \label{tab:serra-cabral-aoi-metrics}
  \begin{tabular}{llcccc}
    \toprule
    Model & Metric & Stage 1 & Stage 2 & Stage 3 & Stage 4 \\
    \midrule
    Logistic Regression & AUC-PR & 0.027 & 0.028 & 0.027 & 0.049 \\
     & AUC-ROC & 0.744 & 0.758 & 0.751 & 0.781 \\
     & Precision & 0.018 & 0.017 & 0.017 & 0.047 \\
     & Recall & 0.832 & 0.863 & 0.847 & 0.076 \\
    \addlinespace
    Random Forest & AUC-PR & 0.061 & 0.135 & 0.262 & 0.100 \\
     & AUC-ROC & 0.836 & 0.919 & 0.955 & 0.893 \\
     & Precision & 0.130 & 0.162 & 0.250 & 0.134 \\
     & Recall & 0.080 & 0.233 & 0.286 & 0.324 \\
    \addlinespace
    XGBoost & AUC-PR & 0.032 & 0.036 & 0.032 & 0.061 \\
     & AUC-ROC & 0.775 & 0.791 & 0.783 & 0.843 \\
     & Precision & 0.017 & 0.017 & 0.018 & 0.018 \\
     & Recall & 0.897 & 0.908 & 0.901 & 0.947 \\
    \bottomrule
  \end{tabular}
\end{table}
\FloatBarrier

\begin{figure}[!htbp]
  \centering
  \includegraphics[width=\linewidth]{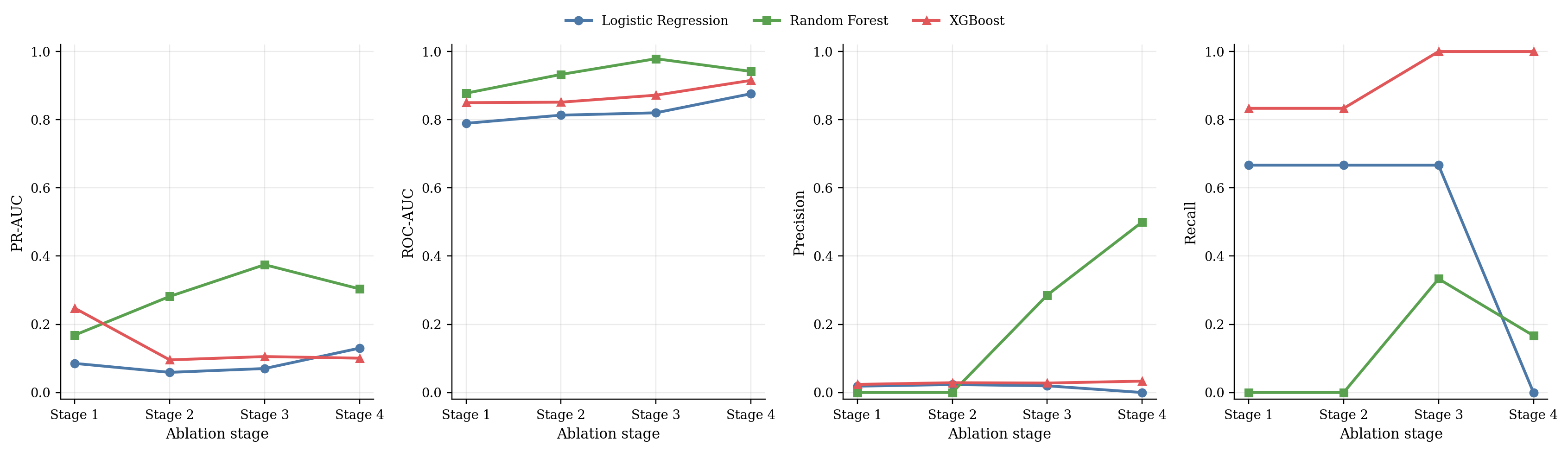}
  \caption{\footnotesize Pau Furado AOI metrics by stage and model (6 positives, 600 pseudo absences).}
  \label{fig:pau-furado-aoi-trends}
\end{figure}

Pau Furado contained only 6 positives and 600 pseudo absences, so its metrics are reported as a small-sample stress test with a low-positive warning.
Within this AOI, the highest Stage 4 AUC-PR was obtained by Random Forest (0.304), followed by Logistic Regression (0.130) and XGBoost (0.101) (Figure~\ref{fig:pau-furado-aoi-trends}).
Across all stages in Pau Furado, Random Forest reached its highest AUC-PR in Stage 3 (0.375), while Stage 4 retained the strongest AUC-PR among the three Stage 4 models.

\FloatBarrier

\begin{table}[!htbp]
  \centering
  \scriptsize
  \caption{Pau Furado AOI test metrics by ablation stage and model.}
  \label{tab:pau-furado-aoi-metrics}
  \begin{tabular}{llcccc}
    \toprule
    Model & Metric & Stage 1 & Stage 2 & Stage 3 & Stage 4 \\
    \midrule
    Logistic Regression & AUC-PR & 0.085 & 0.059 & 0.070 & 0.130 \\
     & AUC-ROC & 0.789 & 0.813 & 0.820 & 0.876 \\
     & Precision & 0.019 & 0.023 & 0.020 & 0.000 \\
     & Recall & 0.667 & 0.667 & 0.667 & 0.000 \\
    \addlinespace
    Random Forest & AUC-PR & 0.168 & 0.282 & 0.375 & 0.304 \\
     & AUC-ROC & 0.878 & 0.933 & 0.979 & 0.942 \\
     & Precision & 0.000 & 0.000 & 0.286 & 0.500 \\
     & Recall & 0.000 & 0.000 & 0.333 & 0.167 \\
    \addlinespace
    XGBoost & AUC-PR & 0.247 & 0.096 & 0.105 & 0.101 \\
     & AUC-ROC & 0.850 & 0.851 & 0.872 & 0.915 \\
     & Precision & 0.024 & 0.029 & 0.028 & 0.033 \\
     & Recall & 0.833 & 0.833 & 1.000 & 1.000 \\
    \bottomrule
  \end{tabular}
\end{table}
\FloatBarrier

\begin{figure}[!htbp]
  \centering
  \begin{minipage}[t]{0.32\linewidth}
    \centering
    \includegraphics[width=\linewidth]{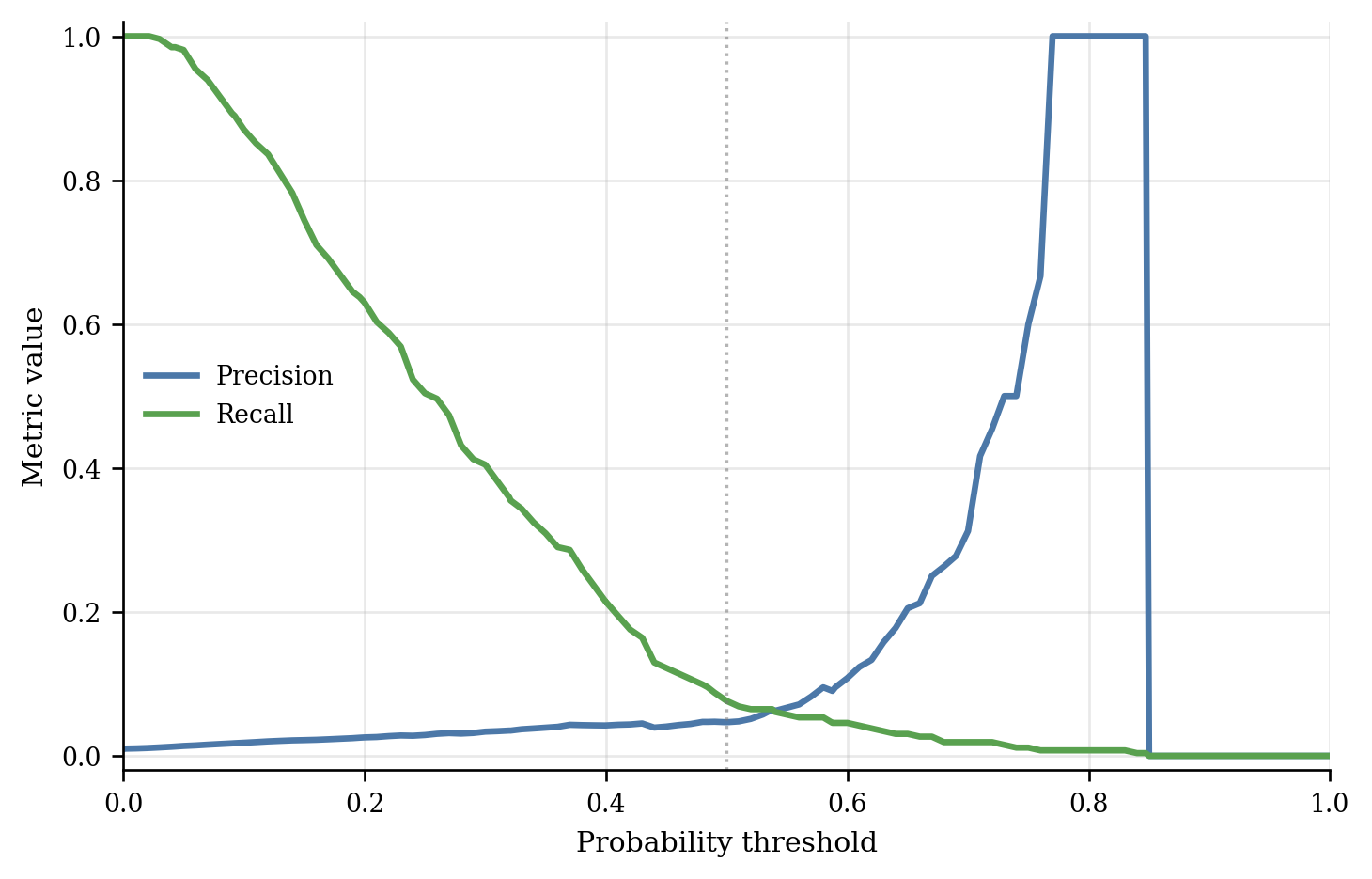}
    \scriptsize \textbf{(a)} Serra do Cabral, Logistic Regression
  \end{minipage}\hfill
  \begin{minipage}[t]{0.32\linewidth}
    \centering
    \includegraphics[width=\linewidth]{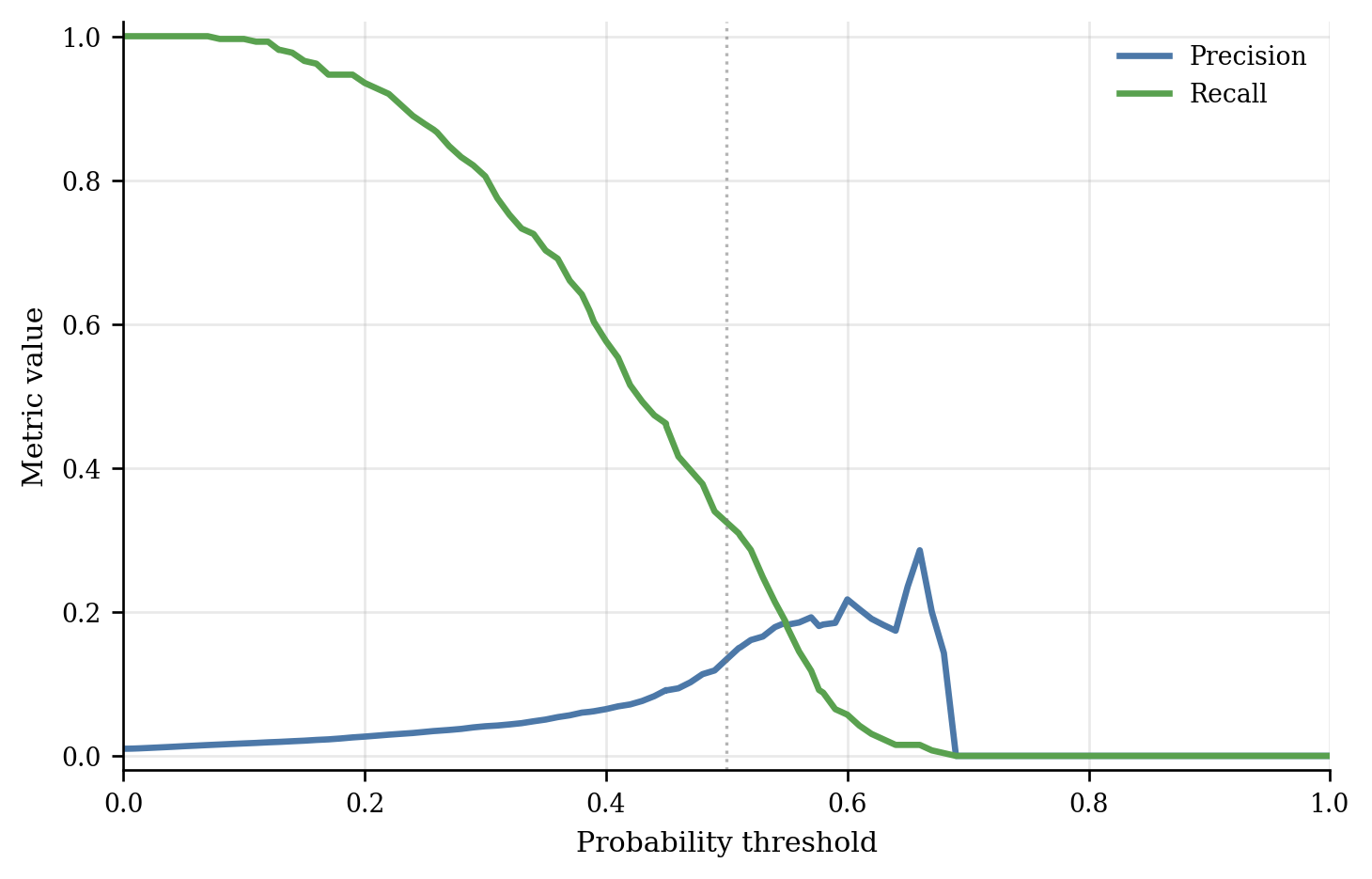}
    \scriptsize \textbf{(b)} Serra do Cabral, Random Forest
  \end{minipage}\hfill
  \begin{minipage}[t]{0.32\linewidth}
    \centering
    \includegraphics[width=\linewidth]{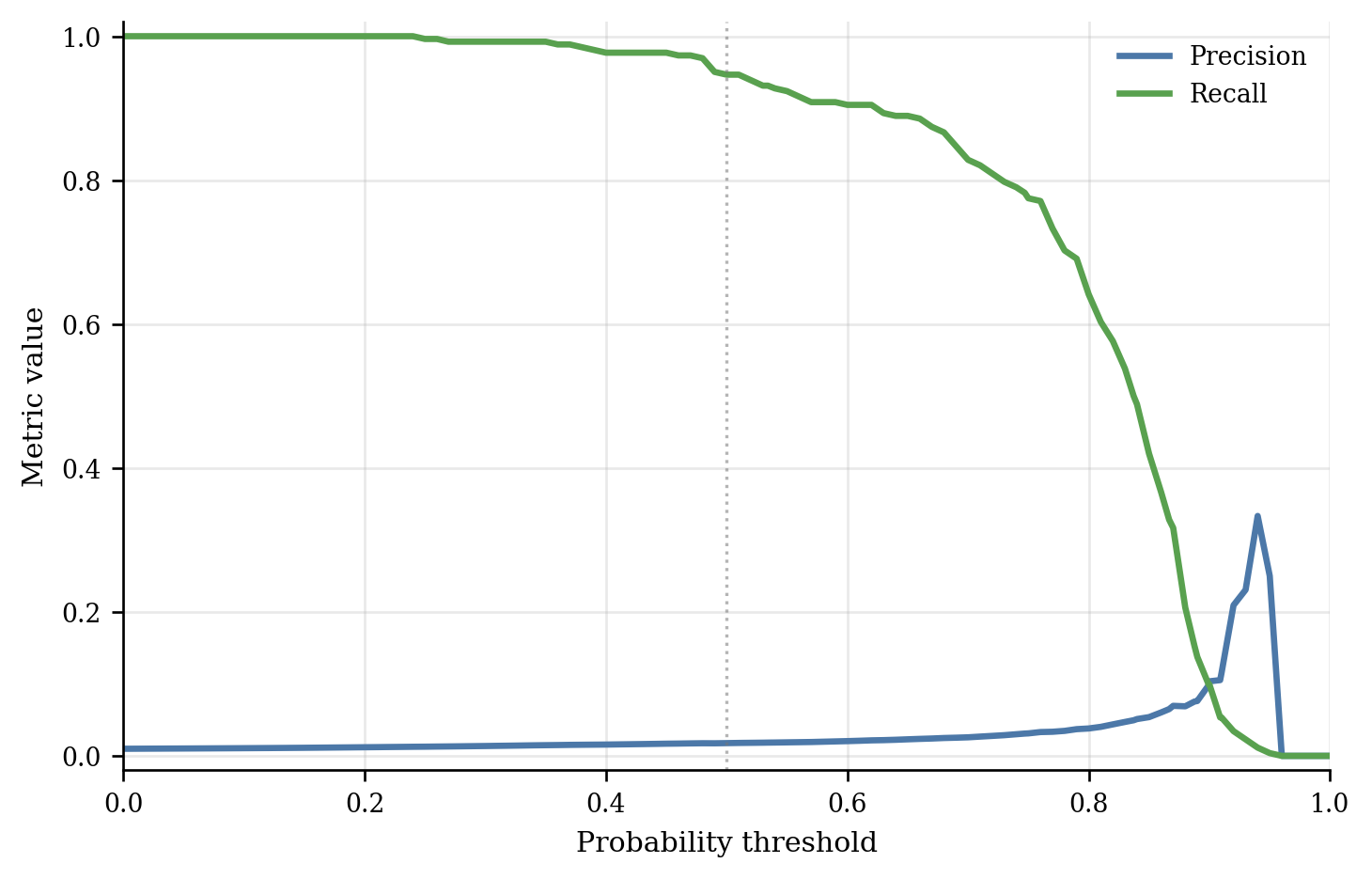}
    \scriptsize \textbf{(c)} Serra do Cabral, XGBoost
  \end{minipage}

  \vspace{0.35em}

  \begin{minipage}[t]{0.32\linewidth}
    \centering
    \includegraphics[width=\linewidth]{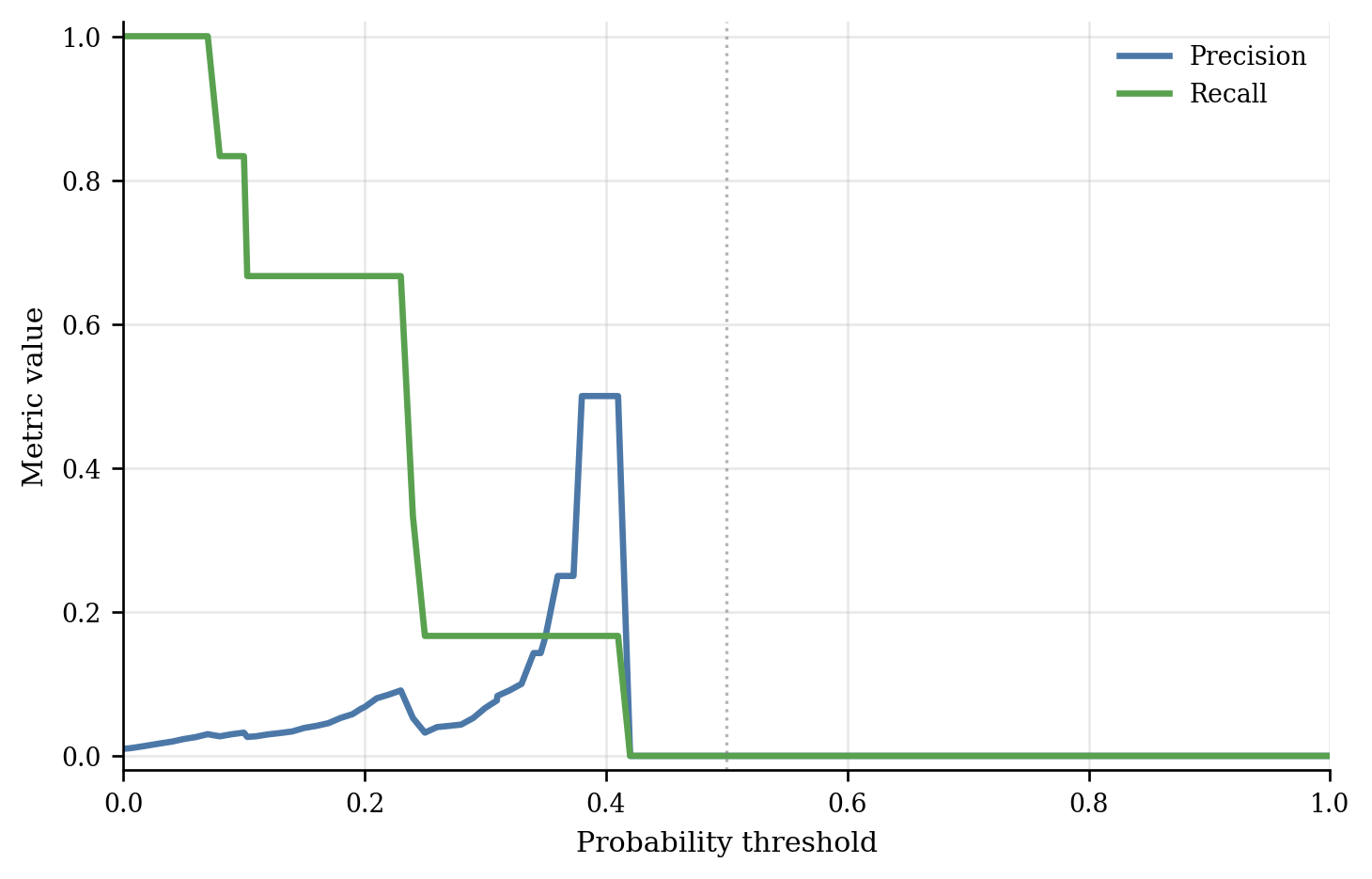}
    \scriptsize \textbf{(d)} Pau Furado, Logistic Regression
  \end{minipage}\hfill
  \begin{minipage}[t]{0.32\linewidth}
    \centering
    \includegraphics[width=\linewidth]{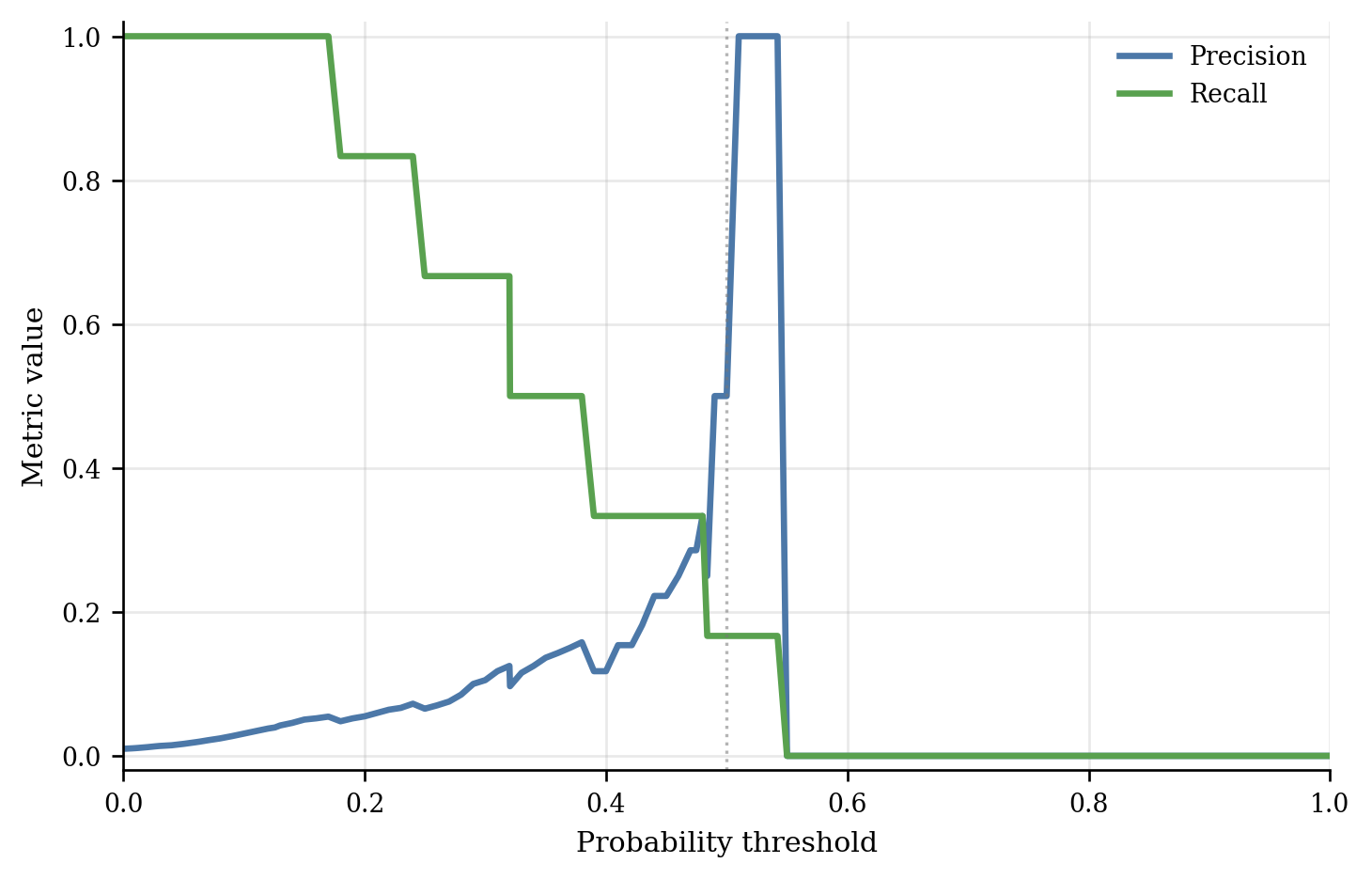}
    \scriptsize \textbf{(e)} Pau Furado, Random Forest
  \end{minipage}\hfill
  \begin{minipage}[t]{0.32\linewidth}
    \centering
    \includegraphics[width=\linewidth]{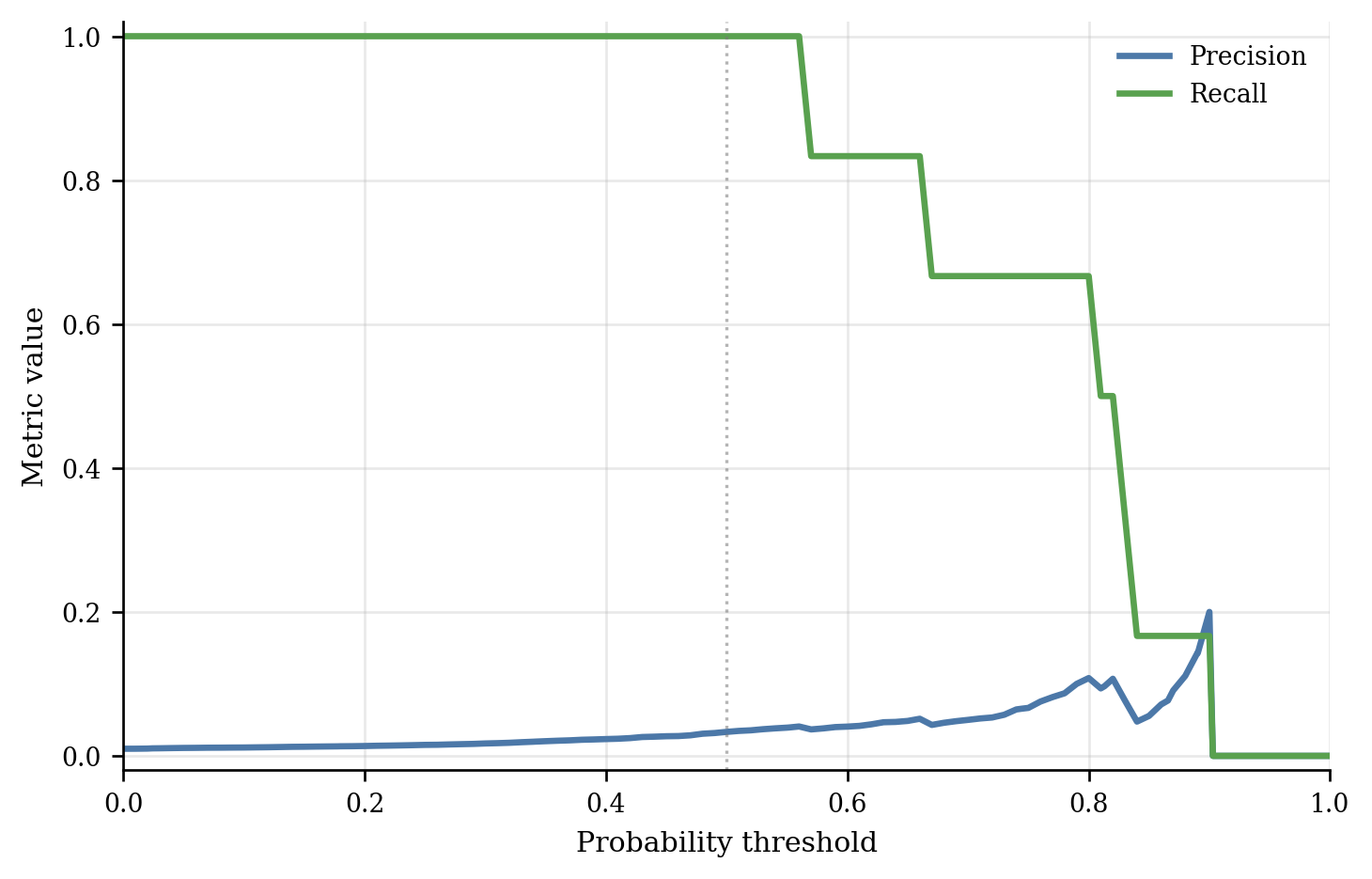}
    \scriptsize \textbf{(f)} Pau Furado, XGBoost
  \end{minipage}
  \caption{\footnotesize Stage 4 AOI Precision and Recall across probability-score thresholds. Top row: Serra do Cabral. Bottom row: Pau Furado.}
  \label{fig:stage4-aoi-threshold-sensitivity}
\end{figure}

Figure~\ref{fig:stage4-aoi-threshold-sensitivity} expands these fixed-threshold readings across the full probability-score threshold range for both AOIs.
The Serra do Cabral row is the more stable transfer diagnostic, while the Pau Furado row should be read as low-support stress-test behavior.

\FloatBarrier

\subsection{AOI-specific Stage 4 explanations}

\begin{figure}[!htbp]
  \centering
  \begin{minipage}[t]{0.45\linewidth}
    \centering
    \includegraphics[width=\linewidth]{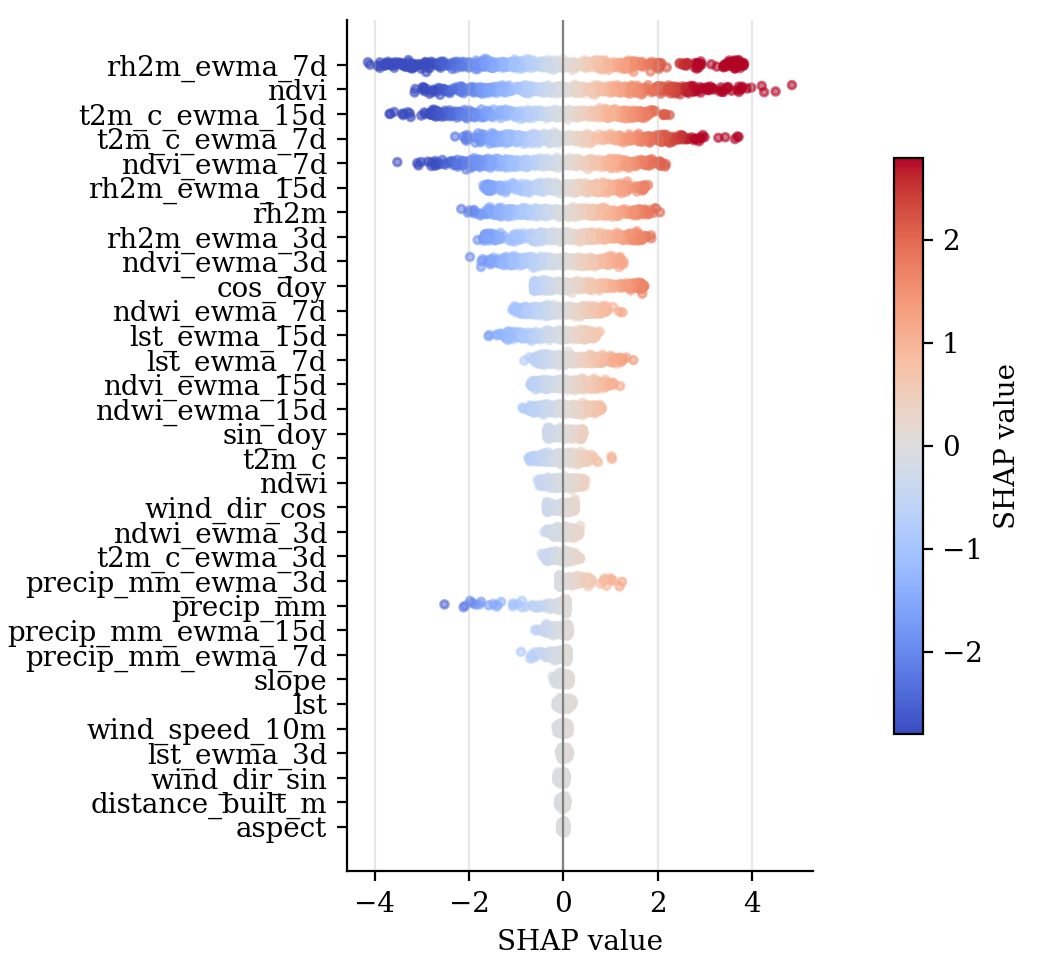}
    \scriptsize \textbf{(a)} Logistic Regression
  \end{minipage}\hfill
  \begin{minipage}[t]{0.45\linewidth}
    \centering
    \includegraphics[width=\linewidth]{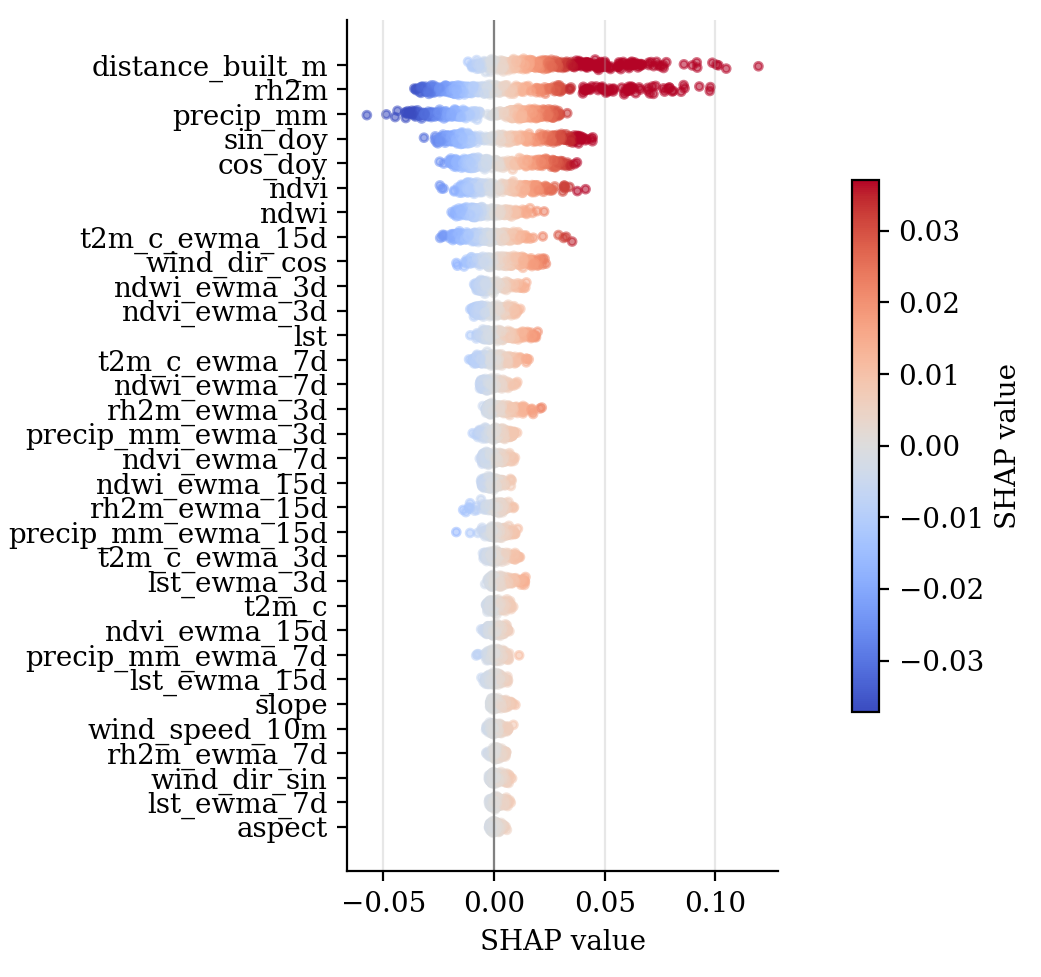}
    \scriptsize \textbf{(b)} Random Forest
  \end{minipage}
  \vspace{0.2em}

  \begin{minipage}[t]{0.45\linewidth}
    \centering
    \includegraphics[width=\linewidth]{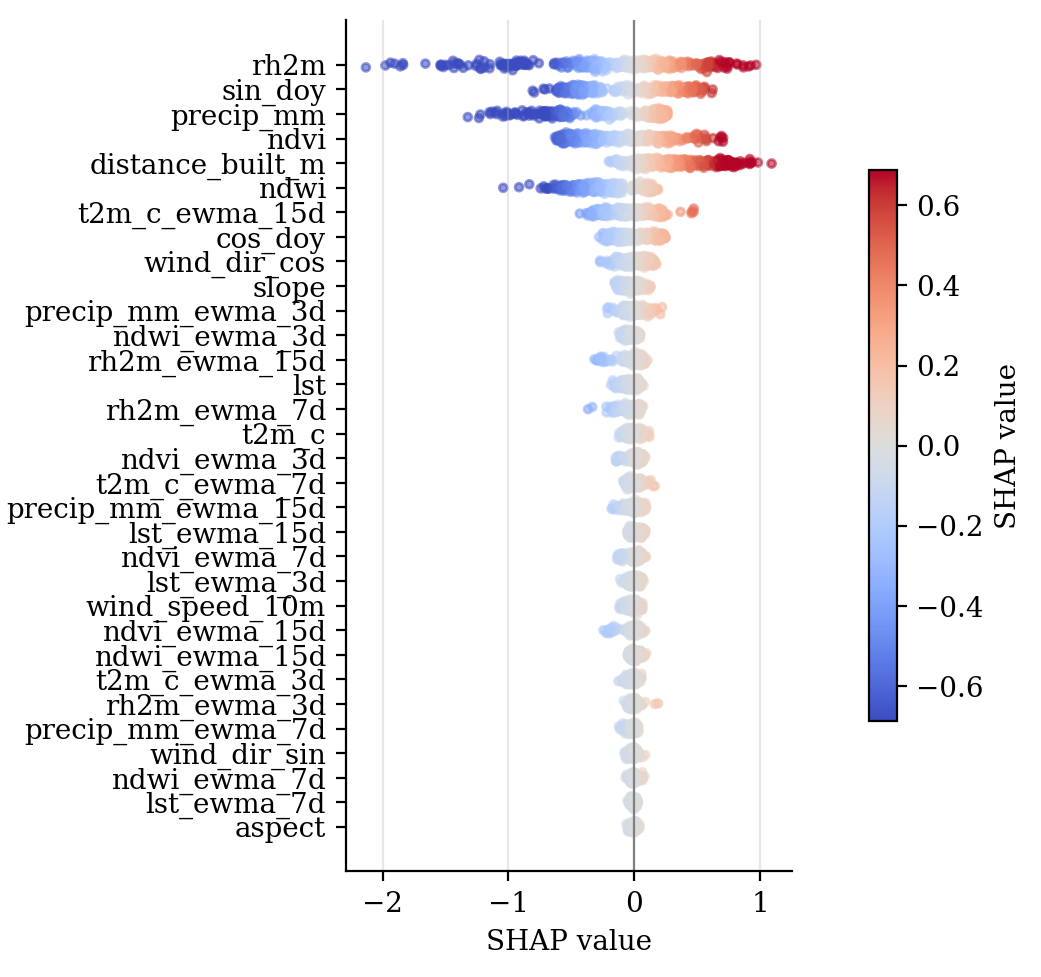}
    \scriptsize \textbf{(c)} XGBoost
  \end{minipage}
  \caption{\footnotesize Stage 4 Pau Furado AOI SHAP summaries for (\textbf{a}) Logistic Regression, (\textbf{b}) Random Forest, and (\textbf{c}) XGBoost.}
  \label{fig:stage4-pau-furado-shap}
\end{figure}

\begin{figure}[!htbp]
  \centering
  \begin{minipage}[t]{0.45\linewidth}
    \centering
    \includegraphics[width=\linewidth]{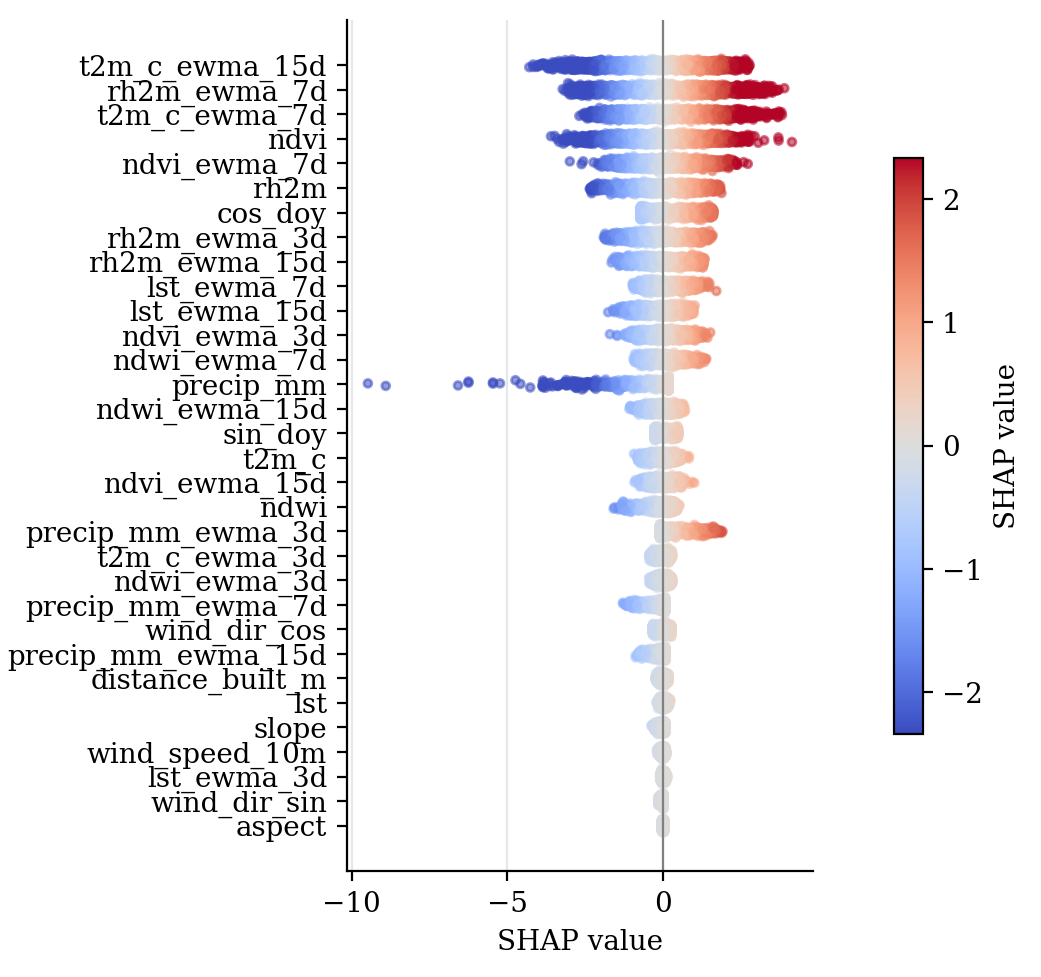}
    \scriptsize \textbf{(a)} Logistic Regression
  \end{minipage}\hfill
  \begin{minipage}[t]{0.45\linewidth}
    \centering
    \includegraphics[width=\linewidth]{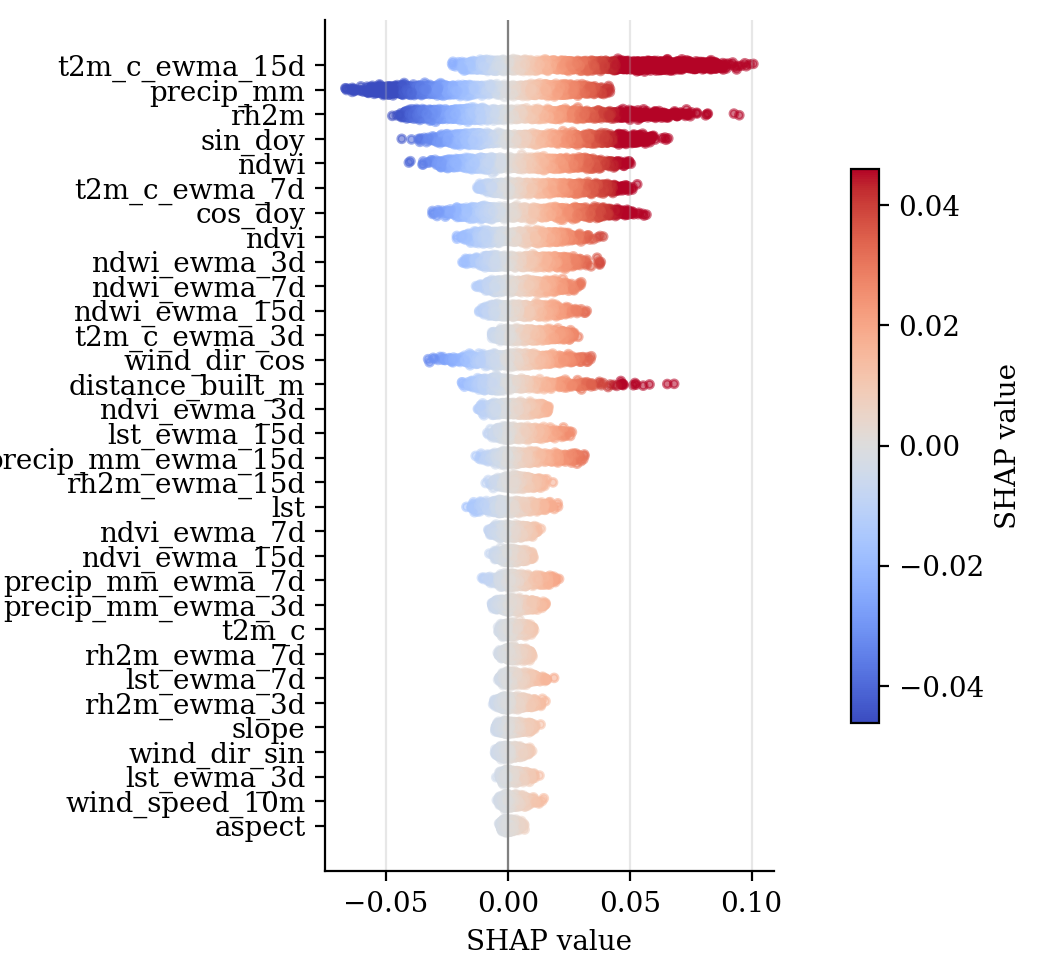}
    \scriptsize \textbf{(b)} Random Forest
  \end{minipage}
  \vspace{0.2em}

  \begin{minipage}[t]{0.45\linewidth}
    \centering
    \includegraphics[width=\linewidth]{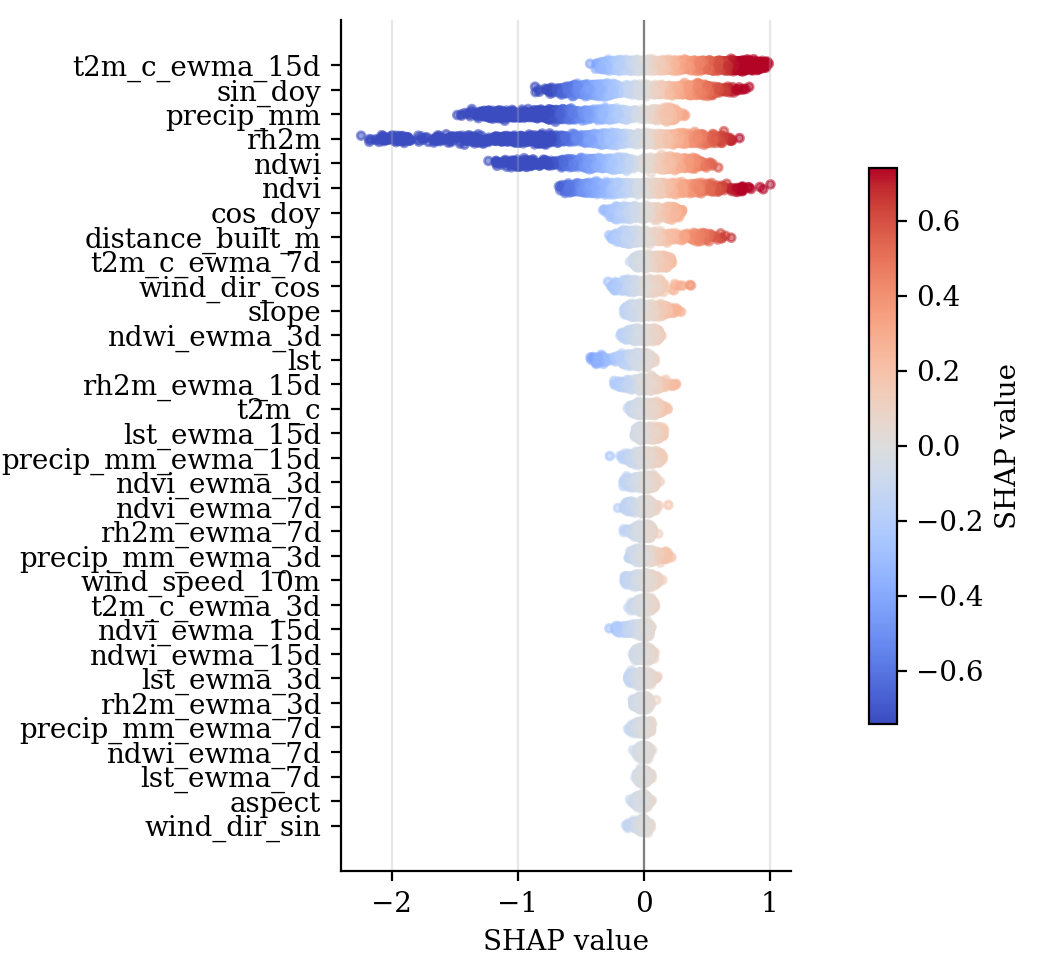}
    \scriptsize \textbf{(c)} XGBoost
  \end{minipage}
  \caption{\footnotesize Stage 4 Serra do Cabral AOI SHAP summaries for (\textbf{a}) Logistic Regression, (\textbf{b}) Random Forest, and (\textbf{c}) XGBoost.}
  \label{fig:stage4-serra-cabral-shap}
\end{figure}

Figures~\ref{fig:stage4-pau-furado-shap} and~\ref{fig:stage4-serra-cabral-shap} show Stage 4 SHAP summaries on the held-out AOI test rows.

\FloatBarrier

For Pau Furado, the small positive support makes the SHAP plots primarily diagnostic of model behavior on the sampled AOI point set.
For Serra do Cabral, the larger test set supports a more stable comparison of AOI-specific explanatory profiles.
Random Forest produced the highest Stage 4 AOI AUC-PR, while XGBoost produced the highest Stage 4 recall at the 0.5 threshold.

\FloatBarrier

\subsection{Operational-style retrospective maps}

The final output is a retrospective diagnostic simulation over dense 500 m AOI grids.
The main text reports one high-fire daily heatmap for each AOI.
The selected examples are Pau Furado on 18 September 2019 and Serra do Cabral on 13 September 2024.
Both figures compare the three Stage 4 model families on the same AOI grid and date, with CU and buffer-zone boundaries overlaid and observed BDQueimadas detections marked on the corresponding map.
These maps connect sampled point tests to continuous spatial score surfaces and were not used for tuning or model selection.
In both examples, XGBoost assigns high scores across a broader share of the mapped surface than Logistic Regression or Random Forest.
This saturation is a warning-volume diagnostic: high recall at sampled points can correspond to excessive mapped alert area.

\begin{figure}[!htbp]
  \centering
  \includegraphics[width=\linewidth]{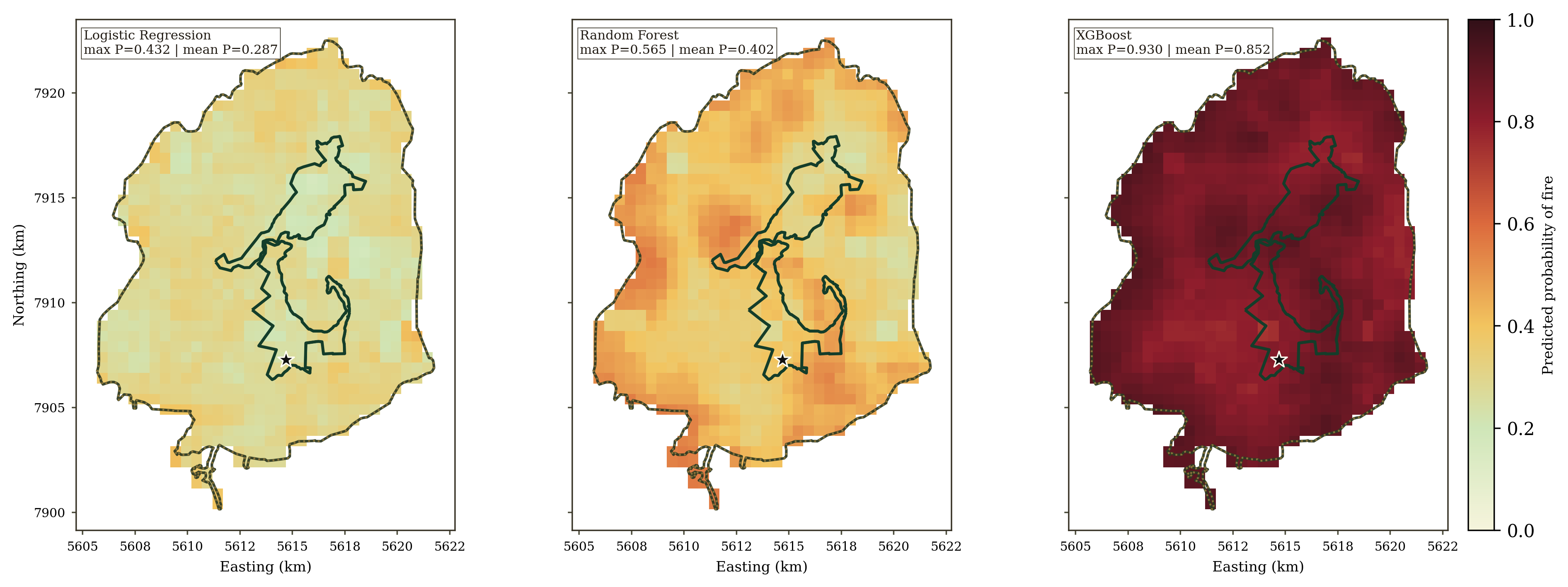}
  \caption{\footnotesize Pau Furado Stage 4 operational-style diagnostic heatmap for 18 September 2019. Panels share the same grid and date.}
  \label{fig:pau-furado-operational-heatmap-20190918}
\end{figure}
\FloatBarrier

\begin{figure}[!htbp]
  \centering
  \includegraphics[width=\linewidth]{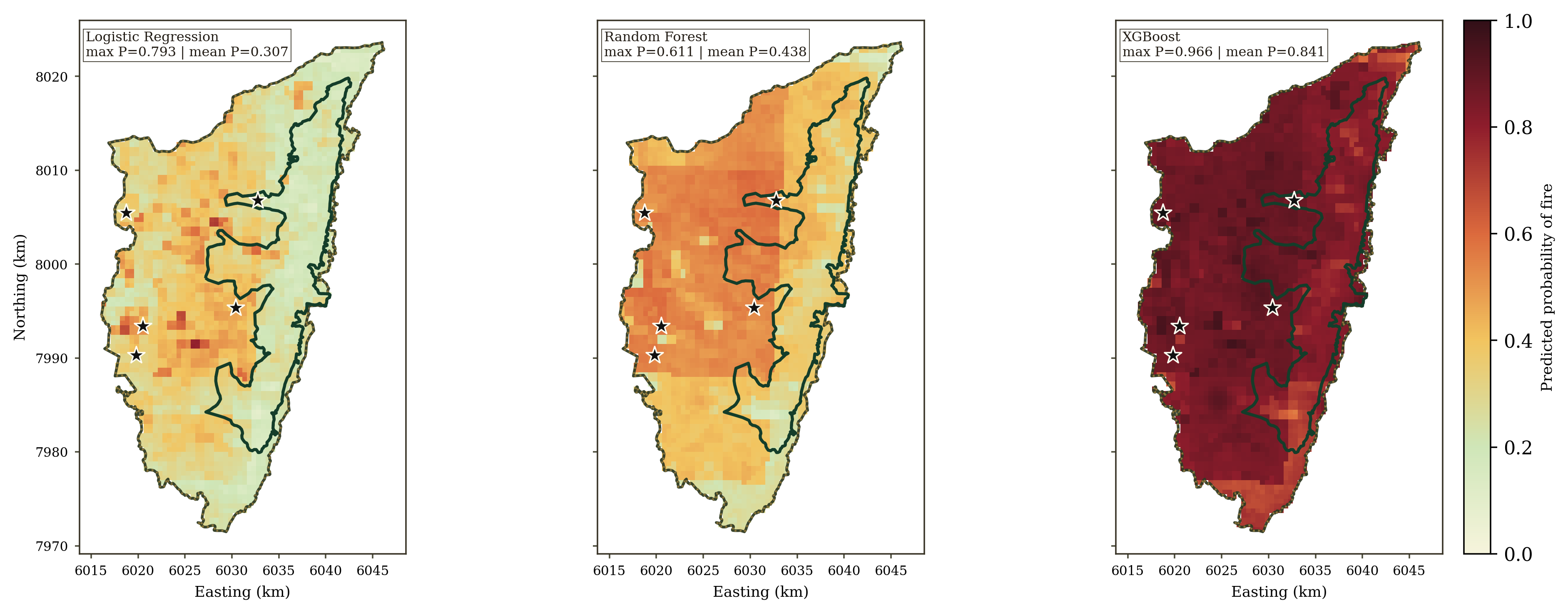}
  \caption{\footnotesize Serra do Cabral Stage 4 operational-style diagnostic heatmap for 13 September 2024. Panels share the same grid and date.}
  \label{fig:serra-cabral-operational-heatmap-20240913}
\end{figure}
\FloatBarrier

\section{Discussion}

\label{sec:discussion}

The results support four main findings.
First, feature fusion improves temporal validation in the Cerrado-MG training domain when atmospheric, surface, static spatial, and temporal-memory covariates are added in a row-aligned sequence.
Second, transfer to held-out CU and buffer-zone surfaces is substantially harder, especially under the 1:100 AOI prevalence design recommended by rare-event evaluation guidance~\citep{phelps2021guidelines}.
Third, Random Forest provides the most balanced AOI behavior in this study, whereas XGBoost tends toward high recall and larger warning volume.
Fourth, the present models are diagnostic baselines; prospective forecasting requires feature availability before decision time, calibration, and prospective validation.

\subsection{Value of feature-level covariate fusion}

The monotonic Stage 1 to Stage 4 temporal-validation AUC-PR gains indicate that the added covariate families carry complementary ranking information.
Because the ablation tables preserve the same rows, dates, labels, and spatial units across stages, the gains reported in Section~\ref{sec:results} are not artifacts of sample replacement.

The daily point-date formulation retains learnable structure after land-cover filtering, event merging, and dry-season restriction.
The causal temporal-memory features add information beyond instantaneous same-day atmospheric and surface conditions.
This pattern matches the logic of operational fire-risk indices such as Risco de Fogo, where recent rainfall history is central, and extends it to heat, humidity, precipitation, greenness, water stress, and surface temperature~\citep{setzer2019riscofogo}.

The result is a ranking improvement, not a deployment claim.
A higher temporal-validation AUC-PR does not guarantee stable transfer to a new CU, calibrated probability scores at deployment prevalence, or acceptable warning burden for managers.

\subsection{AOI transfer and deployment prevalence}

The AOI results separate temporal generalization from spatial transfer.
Serra do Cabral provides the stronger AOI evidence, and Random Forest remained the most effective AOI-ranking model across the main stages.
The Stage 3 peak indicates that static spatial context and surface state can be useful for transfer into this AOI, while the lower Stage 4 AUC-PR relative to Stage 3 suggests that temporal-memory features may introduce additional variance or site-specific behavior.
Feature families with strong temporal-validation value still require testing under the target spatial and prevalence regime.

In Serra do Cabral, XGBoost produced very high recall at the 0.5 threshold but with low precision, whereas Random Forest produced a more balanced but still false-positive-sensitive profile.
The threshold-sensitivity curves show that model families trace different Precision-Recall paths as the probability-score threshold changes.
For future operational fire management, a high-recall model can be valuable only if the warning volume remains interpretable and actionable.

Pau Furado is best interpreted as a low-support AOI stress test rather than a definitive site-level ranking of algorithms.
Within that limited support, Random Forest achieved the highest Stage 4 AUC-PR among the three models, while its strongest Pau Furado value occurred in Stage 3.
This pattern reinforces the need to report low-positive warnings beside threshold diagnostics.

\subsection{Model behavior and explanatory diagnostics}

Across the triad, Logistic Regression functions as a transparent linear reference, Random Forest provides the most stable AOI compromise between ranking and threshold behavior, and XGBoost remains competitive under temporal validation while becoming more recall-oriented under AOI transfer.
This contrast shows how model family, feature stage, and evaluation regime interact~\citep{cilli2022xai,yang2026forests,bian2024forests,freitas2025triunfo}.

The SHAP summaries are model-behavior diagnostics, not direct ecological causality.
They audit whether fitted scores rely on physically plausible drivers such as heat, humidity, precipitation memory, vegetation condition, surface temperature, terrain, and built-environment proximity, while inheriting the sampling design, feature collinearity, sensor limits, and model class.

\subsection{Operational-style maps as retrospective diagnostics}

The heatmaps show that point-level ranking does not by itself guarantee useful spatial behavior.
They are retrospective diagnostics, excluded from tuning and model selection, that connect sampled AOI tests to the continuous surface managers would inspect.
In the selected examples, XGBoost produced the broadest high-score surfaces, reinforcing the warning-volume issue implied by its high-recall AOI profile.
A future prospective study should formalize this visual check through temporal mirror plots, daily alert-volume summaries, calibration curves, feature-availability audits, and season-by-season validation.

\subsection{Limitations and future directions}

BDQueimadas active-fire detections represent the fire signal observed by orbital sensors, not a complete census of all fire events.
Cloud cover, satellite overpass timing, thermal contrast, sub-pixel fire size, and repeated detections can affect positive labels~\citep{giglio2003enhanced,giglio2016collection6,hantson2013strengths}.
The current preprocessing reduces duplicate detections and keeps the reference satellite family consistent; the next round should compare these labels with burned-area products, park incident records, and sensor-confidence filters.

The pseudo-absence design controls season, land cover, distance from same-day positives, and spatial hold-out rules, following pseudo-absence practice in ecological modeling~\citep{barbetmassin2012pseudo}.
Because sampled negatives remain background cases, future sensitivity analysis should test prevalence ratios, distance thresholds, and sampling domains while keeping the 1:100 AOI test as the deployment-prevalence reference.

The two held-out AOIs are legally and operationally meaningful protected-area surfaces, but they do not represent the full diversity of Cerrado CUs, land-tenure contexts, road networks, agricultural frontiers, and fire-management regimes.
A stronger transfer study should add more CUs, more years, and broader buffer-zone contexts while preserving the separation between training geography and independent AOI testing.

Stage 3 represents human-access context through distance to built-up areas.
Ignition pressure in the Cerrado would be described more directly by distance to roads, rural settlements, urban centers, farms or agricultural edges, pasture mosaics, CU boundaries, road density, and recent land-cover transition metrics.
These variables would better represent the human-ignition mechanisms discussed in Brazilian and international wildfire literature~\citep{jain2020review,hoffmann2020fire,freitas2025triunfo}.

Stage 4 uses causal EWMA features as an auditable temporal-memory baseline, but the 3, 7, and 15 day grid is intentionally short.
It does not represent seasonal drought memory, such as the slower Drought Code component of the Canadian FWI System or the 120-day rainfall history in the INPE Risco de Fogo index~\citep{vanwagner1987fwi,setzer2019riscofogo}.
Future experiments should compare EWMAs with denser temporal representations, including convolutional time-series transforms, state-space or SARIMAX-style summaries, temporal clustering of pre-fire trajectories, sequence-aware neural networks, and hybrid representations that preserve both short dry-spell intensity and longer seasonal accumulation.

The current same-day covariate design is retrospective by construction.
Stages 1 to 3 use environmental conditions from the target date, and Stage 4 adds causal memory without replacing those same-day inputs.
Therefore, the study supports daily active-fire detection classification and ranking, not next-day operational forecasting.
Prospective forecasting would require predictors known before the decision time, such as D-1 meteorological fields, forecast products, or explicitly lagged remote-sensing summaries, followed by calibration and prospective validation.

Future model development can broaden beyond the three-classifier baseline once the training set expands to more CUs, denser gridded histories, or image-like spatiotemporal tensors~\citep{jain2020review}.
Probabilistic nonlinear models could quantify uncertainty in rare high-risk areas, and an ensemble decision layer could combine calibrated scores, uncertainty estimates, alert-volume limits, and feature-attribution consistency under predefined operational thresholds.

\section{Conclusions and Final Remarks}

This study presents a reproducible retrospective daily active-fire detection benchmark for Cerrado CUs in Minas Gerais, combining BDQueimadas active-fire labels, constrained pseudo absences, row-aligned feature ablation, temporal validation, independent AOI testing, SHAP diagnostics, and retrospective heatmaps.
The four-stage ablation shows that causal temporal-memory augmentation adds measurable ranking information under five-fold temporal cross-validation, with Stage 4 producing the highest mean AUC-PR and AUC-ROC for all three model families.
Transfer performance decreases under stricter 1:100 prevalence and spatial hold-out conditions, especially where positive support is small, but Random Forest indicates that useful AOI ranking structure remains detectable.

The main contribution of the work is therefore a transparent experimental baseline for asking which covariate families, model classes, and validation designs are credible for CU-scale daily active-fire detection ranking in the Cerrado.
Because the present stages include same-day environmental covariates, they should not be read as a completed operational forecast.
Future versions should enforce predictor availability before the decision time through D-1 covariates, forecast products, or explicitly lagged temporal windows, while also expanding anthropogenic ignition proxies, comparing richer temporal-memory representations, and testing calibrated decision rules under fire-season monitoring.
These extensions can build on the same row-aligned ablation contract and AOI testing design introduced here.

\section*{Acknowledgments}

The author acknowledges the auxiliary use of generative AI tools during the preparation of this manuscript and the computational workflow, including support for code drafting, debugging, documentation, LaTeX editing, and language revision. These tools were used only as assistive tools and were not treated as sources of scientific evidence, methodological authority, or autonomous analysis.

\bibliography{references}

\end{document}